# Cost Analysis for Import and Export Using an Abstract Machine

cs-24-dat-8-02
Benjamin Bennetzen, Daniel Vang Kleist, Emilie Sonne Steinmann, Loke Walsted, Nikolaj Rossander Kristensen, and Peter Buus Steffensen

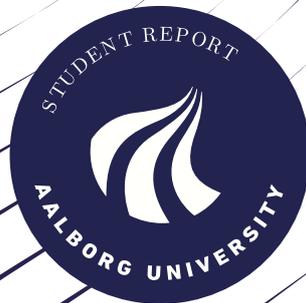

**Abstract:**

This paper presents the syntax and reduction rules for an abstract machine based on the JavaScript XML language.

We incorporate the notion of cost into our reduction rules, and create a type system that over-approximate this cost. This over-approximation results in an equation that may contain unknowns originating from while loops.

We conclude with a formal proof of soundness of the type system for our abstract machine, demonstrating that it over-approximates the cost of any terminating program. An implementation of the type system, constraint gathering, and the abstract machine is also presented.

# Contents

# 1 Introduction

When working in web design, many components used by the programmer or designer are imported from other sources. These dependencies can have their own dependencies, and thereby vary in size. Therefore, the computational cost of inclusion of a dependency also varies. This becomes an issue when a program, such as a web-based design tool, has to operate at 60 frames pr. second for an optimal user experience. This smooth experience could be challenged when including a new element that is not already imported and is computationally expensive. Analyzing the cost of such an import could help prove what impact it has on the experience. In this paper we want to examine a method that could be used for cost analysis.

A lot of different work has been done tackling similar problems. Avanzini and Lago [1] introduces a method for complexity analysis for a functional programming language using a type system. Hoffmann et al. [2] gives a system for getting the worst-case resource bound and Vasconcelos [3] gives a method for program analysis in regards to space cost.

One method that is recurring in some of these analyses is using a type system with sized types. Sized types as introduced by Hughes et al. [4] were used for proving properties such as liveness in reactive systems. This was later extended by Hughes and Pareto [5] to approximate stack and heap cost. Sized types are useful as they can be used to carry information about objects. In this paper, we will use sized types to carry information about the computational cost of files.

We will show an analysis method for a subset of JavaScript XML inspired by Baillot and Ghyselen [6]. The paper examines a method for finding the time complexity for a process in the $\pi$-calculus, by introducing semantics using a *tick*-notation to denote operations with an impact on the time complexity, as well as a type system using sized types. An interesting method of carrying out our analysis would be to make use of an abstract machine. Diehl et al. [7] give an overview of different abstract machines and how they are used. We will introduce an abstract machine similar to the one introduced by Montenegro et al. [8]. Our abstract machine is in the style of Landin's SECD machine [9]. We will annotate its transitions with a cost using the *tick*-notation. In our cost model, we are interested in import and export so we will use the *tick* to mark reductions that will have an impact on this. We will also use *tick* to notify writing to memory, to show that programs without import and export also might have a substantial cost.

To better illustrate the use of our abstract machine and its integrated type system, we will use the following JavaScript XML in Listing 1 and Listing 2, throughout the paper to demonstrate the process from written code to the abstract machine and the type system. Note that these are written such that they can be parsed by our implementation, and as such are not syntactically correct JavaScript XML programs.

```
1  // FILENAME: /simpleWhile.jsx
2  let x = 3;
3  while (x) {
4    x = - x 1;
5  };
6  export x;
```

Listing 1: /simpleWhile.jsx exporting the variable x

```
1   // FILENAME: /main.jsx
2   import { x } from "/simpleWhile.jsx";
3
4   let y = 0;
5
6   let func = <prop>
7     y = (+ prop 2);
8   </>;
9
10  comp func (x);
```

Listing 2: /main.jsx importing `x`, and using it as an argument to a component (`func`)

Listing 1 is a simple program that introduces loops. This is important, as it is not always possible to know how many times a loop will execute, meaning it could potentially have an infinite cost.

Listing 2 is an example of a simple program that depends on another file. This could potentially be a file with an incredibly large cost, so even though the file initially seemed cheap to execute, the actual cost of doing it might become incredibly large. Therefore, we will introduce a type system to estimate the cost of such programs.

This paper continues with the following structure: In Section 2, we introduce an abstract machine for a subset of the JavaScript XML language. The abstract machine is used to simulate runs of JavaScript XML files, such that their cost can be approximated. In Section 3, we present a type system that gives an over-approximation of the cost of evaluating JavaScript XML files. In Section 4, we prove that our type system is sound, meaning that it will always over-approximate the cost of actually executing the file in our abstract machine. In Section 5.1, we introduce a way to infer the type using the type system, as well as an implementation. Finally, we discuss the results as well as future work in Section 6.

## 2 Abstract Machine

### 2.1 Abstract Syntax

We define an abstract syntax as seen on Figure 1 for the subset of JavaScript XML which we focus on in this project. We have chosen these specific constructs, because it is a small Turing-complete language extended with the language constructs that we estimate to have the highest impact on the execution cost. In the formation rules, `Components` refer to components as seen in web-frameworks, such as React and Vue. In our context, components can be thought of as functions without a return value. While there is no syntactic way of creating a record, projection still exists, as importing all variables from a file implicitly binds them to a record.

We define $id$ to be the set of all the possible identifiers over some alphabet $\Sigma$. We use $e$ to denote the set of all possible expressions and $S$ to denote the set of all possible statements.

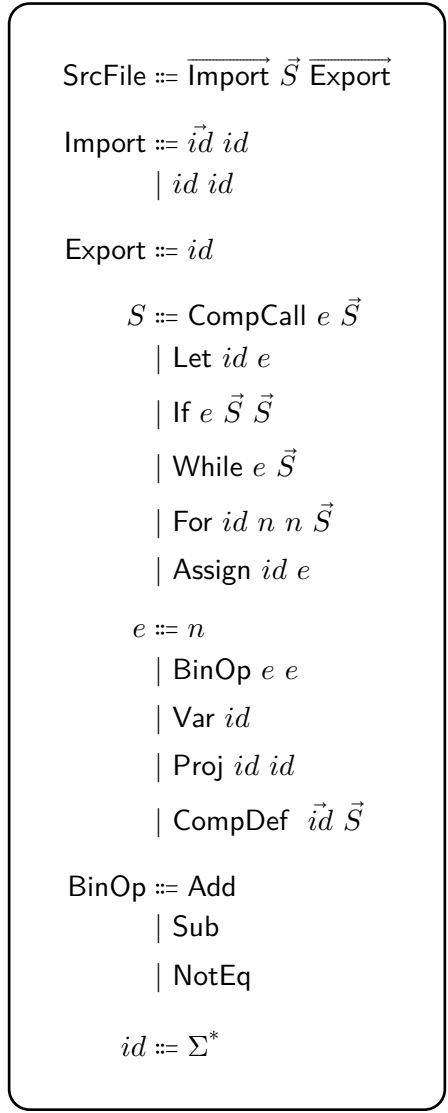

Figure 1: Abstract syntax of JavaScript XML subset

In Appendix A, Listing 1 and Listing 2 can be seen rewritten using the abstract syntax from Figure 1.

## 2.2 Runtime Syntax

We define a runtime syntax for our abstract machine as seen on Figure 2. These formation rules show what constitutes an instruction stack, a value stack ($vs$), and a scope stack ($ss$) in our abstract machine.

The instruction stack is a stack of instructions, where the formation rules for each instruction is all the possible formations in our abstract syntax in Figure 1, with the addition of some specific runtime constructs, e.g., PopScope. The value stack is a stack of values, which can be integers, objects, or components (represented as a function closure). Lastly, the scope stack is simply a stack of identifiers representing each scope. Together with variable identifiers, the scope identifiers are used to retrieve the value of a given variable in a given scope.

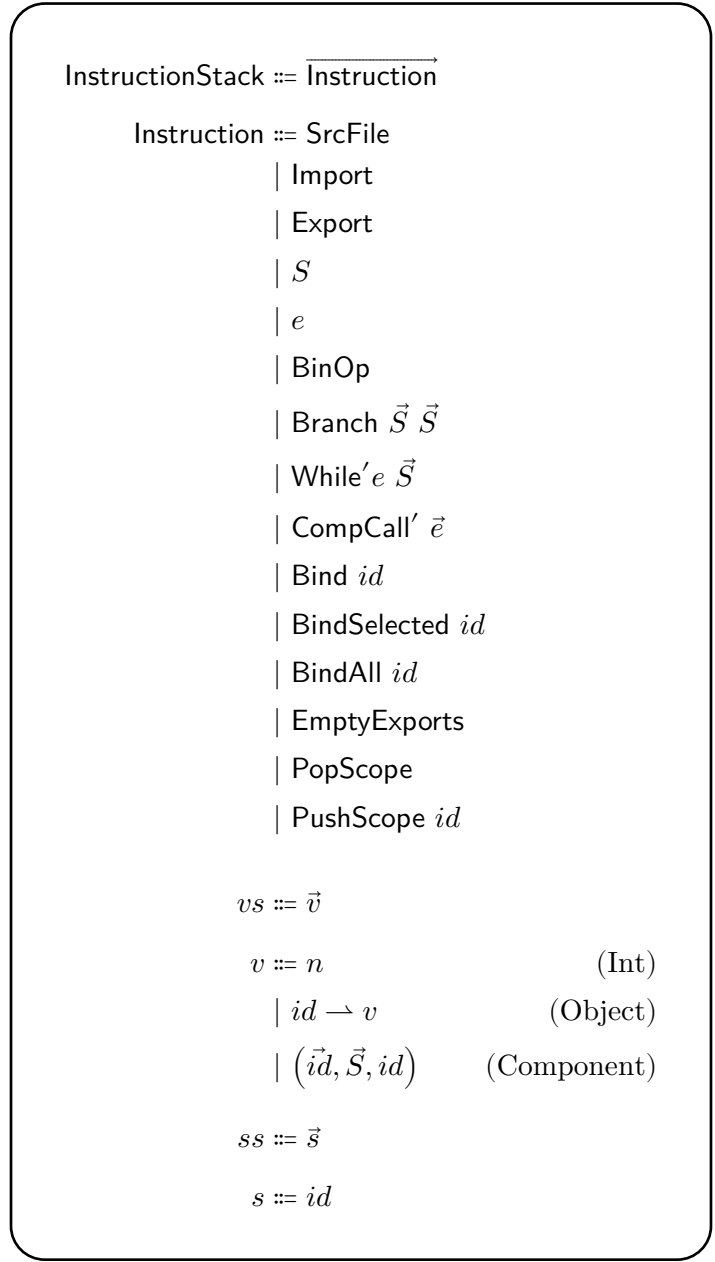

Figure 2: Runtime syntax

## 2.3 Machine Configuration

A machine configuration of the abstract machine follows Definition 1. Here, the instruction stack, value stack ($vs$), and scope stack ($ss$) follows the runtime syntax in Figure 2. $fg$ is a partial function used to retrieve a source file based on its id, $ls$ is a partial function from a pair (scope id, variable id) to the value of the variable in the specified scope, and $es$ is a partial function from $id$ to $v$. The initial state of our abstract machine is defined in Definition 2, and the final state is defined in Definition 3.

**Definition 1** (Machine Configuration): The configuration of the abstract machine is defined as a 6-tuple:

$$\langle \mathsf{InstructionStack}, fg, ls, es, vs, ss \rangle$$

where

$$fg : \mathsf{Id} \rightharpoonup \mathsf{SrcFile}$$
$$ls : \mathsf{Id} \times \mathsf{Id} \rightharpoonup v$$
$$es : \mathsf{Id} \rightharpoonup v$$

**Definition 2** (Initial state): Given a filegetter $fg$ and an entry point $f$, the initial state for the abstract machine is:

$$\langle fg\,(f) : nil, fg, [], [], nil, f : nil \rangle$$

where [] denotes any partial function where the domain is the empty set.

**Definition 3** (Final state): The final state for the abstract machine is any state where the InstructionStack is nil:

$$\langle nil, \_, \_, \_, \_, \_ \rangle$$

## 2.4 Reduction rules

The full set of reduction rules for the abstract machine can be seen in Table 3 in Appendix B. Every reduction $\stackrel{c}{\Longrightarrow}$ has an associated cost $c$. As an example, $\stackrel{0}{\Longrightarrow}$ denotes a reduction with a cost of 0. In Table 1, we present a subset of the reduction rules, which will be described here. The rest of the rules are described in Appendix B.

The (R-BinOp1) and (R-BinOp2) rules reduce a binary operation $\mathsf{BinOp}\ e_1 e_2$ by pushing the $\mathsf{BinOp}$ runtime construct and the two expressions on to the instruction stack. When we reach the $\mathsf{BinOp}$ construct on the instruction stack, we know that the result $v_1$ and $v_2$ of the operands must be the on top of the value stack. We continue by pushing the result of the binary operation on $v_1$ and $v_2$ on to the value stack. We let $bc(\mathsf{BinOp})$ denote the cost of the $\mathsf{BinOp}$. Note that binary operators need not have the same cost. For the purpose of this project, all the binary operators in the syntax have a cost of 0.

(R-CompDef) reduces a $\mathsf{CompDef}$ by creating a component value $(\vec{id}, \vec{S}, s)$ from the component definition $\vec{id}, \vec{S}$ and the current scope $s$, and then pushing said value on to the value stack.

The (R-If) rules describe how we reduce an $\mathsf{If}$-statement. It first pushes the $\mathsf{Branch}\ S_1\ S_2$ construct and then the conditional expression $e_1$ on to the instruction stack. Once the $\mathsf{Branch}$ construct is reached in the instruction stack, the value $v$ of the conditional expression must be the top element of the value stack. In the case $v \neq 0$, we push $S_1$ from the true-block on to the instruction stack, following rule (R-IfTrue), and in the case of $v = 0$, we push $S_2$ from the false-block on to the instruction stack, following rule (R-IfFalse). In this paper the branches do not introduce a new scope, as we deemed it outside the scope of this project.

The rules (R-While), (R-WhileTrue), and (R-WhileFalse) work such that whenever a regular While construct is on the instruction stack we push the While' construct and the conditional expression $e$ on to the instruction stack. When we see the While' construct on the instruction stack, the result of the expression, $v$, must be on top of the value stack. In the case $v \neq 0$ we push the body of the while loop, $\vec{S}$, the conditional expression $e$ and another While' onto the stack. In the case that the result of the conditional expression is $v = 0$ we can simply jump to the next instruction on the stack. In this paper we do not introduce a new scope when entering a While as we deemed it outside the scope of this project.

The (R-Let), (R-Assign), and (R-Bind) rules handle variable bindings and assignments. The abstract machine does not differentiate binding and assignment, so in either case the Bind construct and the expression $e$ is pushed to the instruction stack. Once a Bind construct is seen on the instruction stack, we know that the value of the expression, $v$, is on top of the value stack. As such we can extend the locals with a new mapping from the identifier and current scope to the value $ls[(s, id) \mapsto v]$.

(R-ImportSelected) imports only the set of explicitly included variables from a file. This is handled using a BindSelected construct for each of the $id$'s in the import statement on the instruction stack.

(R-For) handles For loops for which we statically know the ranges $n_1$ to $n_2$. Therefore, we can completely unfold the loop on the instruction stack as soon as it appears.

| Name | Rule | Side condition |
|---|---|---|
| (R-BinOp1) | $\langle \text{BinOp } e_1\ e_2 :: rest, fg, ls, es, vs, ss\rangle$ $\stackrel{0}{\Longrightarrow} \langle e_1 :: e_2 :: \text{BinOp} :: rest, fg, ls, es, vs, ss\rangle$ | |
| (R-BinOp2) | $\langle \text{BinOp} :: rest, fg, ls, es, v_2 :: v_1 :: vs, ss\rangle$ $\stackrel{c}{\Longrightarrow} \langle rest, fg, ls, es, v_3 :: vs, ss\rangle$ | $v_3 = \text{BinOp}\ (v_1, v_2)$ $c = bc\ (\text{BinOp})$ |
| (R-CompDef) | $\langle \text{CompDef } \vec{id}\ \vec{S} :: rest, fg, ls, es, vs, s :: ss\rangle$ $\stackrel{0}{\Longrightarrow} \langle rest, fg, ls, es, (\vec{id}, \vec{S}, s) :: vs, s :: ss\rangle$ | |
| (R-If) | $\langle \text{If } e_1\ \vec{S_1}\ \vec{S_2} :: rest, fg, ls, es, vs, ss\rangle$ $\stackrel{0}{\Longrightarrow} \langle e_1 :: \text{Branch } \vec{S_1}\ \vec{S_2} :: rest, fg, ls, es, vs, ss\rangle$ | |
| (R-IfTrue) | $\langle \text{Branch } \vec{S_1}\ \vec{S_2} :: rest, fg, ls, es, v :: vs, ss\rangle$ $\stackrel{0}{\Longrightarrow} \langle \vec{S_1} :: rest, fg, ls, es, vs, ss\rangle$ | $v \neq 0$ |
| (R-IfFalse) | $\langle \text{Branch } \vec{S_1}\ \vec{S_2} :: rest, fg, ls, es, v :: vs, ss\rangle$ $\stackrel{0}{\Longrightarrow} \langle \vec{S_2} :: rest, fg, ls, es, vs, ss\rangle$ | $v = 0$ |
| (R-While) | $\langle \text{While } e\ \vec{S}, fg, ls, es, vs, ss\rangle$ $\stackrel{0}{\Longrightarrow} \langle e :: \text{While}'\ e\ \vec{S} :: rest, fg, ls, es, vs, ss\rangle$ | |
| (R-WhileTrue) | $\langle \text{While}'\ e\ \vec{S} :: rest, fg, ls, es, v :: vs, ss\rangle$ $\stackrel{0}{\Longrightarrow} \langle \vec{S} :: e :: \text{While}'\ e\ \vec{S} :: rest, fg, ls, es, vs, ss\rangle$ | $v \neq 0$ |
| (R-WhileFalse) | $\langle \text{While}'\ e\ \vec{S} :: rest, fg, ls, es, v :: vs, ss\rangle$ $\stackrel{0}{\Longrightarrow} \langle rest, fg, ls, es, vs, ss\rangle$ | $v = 0$ |
| (R-Bind) | $\langle \text{Bind } id :: rest, fg, ls, es, v :: vs, s :: ss\rangle$ $\stackrel{1}{\Longrightarrow} \langle rest, fg, ls[(s, id) \mapsto v], es, vs, s :: ss\rangle$ | |
| (R-ImportSelected) | $\langle \vec{id}\ f :: rest, fg, ls, es, vs, ss\rangle$ $\stackrel{2}{\Longrightarrow} \langle fg(f) :: \text{PopScope} :: \text{BindSelected } id_1 ::$ $... :: \text{BindSelected } id_n :: \text{EmptyExports} ::$ $rest, fg, ls, es, f :: ss\rangle$ | |
| (R-For) | $\langle \text{For } id\ n_1 n_2\ \vec{S} :: rest, fg\ ls, es, vs, ss\rangle$ $\stackrel{0}{\Longrightarrow} \langle \text{Let } id\ n_1 :: S :: ... :: \text{Let } id\ n_i :: S ::$ $... :: \text{Let } id\ n_2 :: S :: rest, fg, ls, es, vs, ss\rangle$ | $n_i = \mathbb{N}^{-1}(\mathbb{N}(n_1) + i)$ $\mathbb{N}(n_1) \leq \mathbb{N}(n_2)$ |

Table 1: Subset of reduction rules

## 3 Type System

### 3.1 Types

We will now introduce a type system to over-approximate the cost of executing a file on the abstract machine. As our language is Turing-complete, it is not possible to find the exact cost, as it is undecidable if a program will even terminate. Therefore, we will only make sure that if a file is well-typed, then we know that the cost of executing a terminating file on the abstract machine is less than or equal to the value of the type of the file.

The type judgements are of the form $\Gamma \vdash e : t$ for expressions and $\Gamma \vdash S : t \triangleright \Gamma'$ for statements, where $\Gamma$ is a partial function from variables to types. $t$ are the types of our type system, which are defined through the formation rules in Figure 3. Imports and exports are typed similarly to statements.

The type system will be used to over-approximate the cost of a given JavaScript XML program. We will be using two different type systems, one for statements, and one for expressions.

$$
\begin{aligned}
t &::= t_{\text{Stm}} \mid t_{\text{Expr}} \\
t_{\text{Stm}} &::= n &&\text{(number)} \\
&\mid x &&\text{(variable)} \\
&\mid t_{\text{Stm}} + t_{\text{Stm}} &&\text{(adddition)} \\
&\mid t_{\text{Stm}} \cdot t_{\text{Stm}} &&\text{(multiplication)} \\
&\mid t_{\text{Stm}} \uparrow t_{\text{Stm}} &&\text{(maximum)} \\
t_{\text{Expr}} &::= t_{\text{Stm}} &&\text{(actual cost)} \\
&\mid t_{\text{Expr}} \rightarrow t &&\text{(component)} \\
&\mid \left\{ id_i : t_{\text{Expr}_i} \right\}_{i \in 1 \ldots n} &&\text{(record)}
\end{aligned}
$$

Figure 3: Formation rules of types

The type of statements is a max-plus-semiring algebra [10]. The type of expressions includes the type of statements, as well as arrow types to type components, and record types to type imported records.

## 3.2 Partial Order on Types and Type Environments

We define equivalence rules between types as seen on Figure 4. These describe the following properties: the multiplication and addition formation rules are commutative. We can reduce constants that are added or multiplied together. The number 0 is an identity element for addition and the number 1 is an identity element for multiplication. Lastly you can always add a variable term as long as the constant is 0. These equivalence rules allow us to reduce the number of inference rules needed to define the partial order between types.

$$t_{\text{Stm}} \cdot t_{\text{Stm}}' \equiv t_{\text{Stm}}' \cdot t_{\text{Stm}}$$
$$t_{\text{Stm}} + t_{\text{Stm}}' \equiv t_{\text{Stm}}' + t_{\text{Stm}}$$
$$n + m \equiv l \quad \text{where } l = \mathbb{N}^{-1}(\mathbb{N}(n) + \mathbb{N}(m))$$
$$n \cdot m \equiv l \quad \text{where } l = \mathbb{N}^{-1}(\mathbb{N}(n) \cdot \mathbb{N}(m))$$
$$t_{\text{Stm}} \equiv t_{\text{Stm}} + 0$$
$$t_{\text{Stm}} \equiv 1 \cdot t_{\text{Stm}}$$
$$t_{\text{Stm}} \equiv t_{\text{Stm}} + 0 \cdot x$$

Figure 4: The equivalence rules of types.

We define the partial order on types through the definition of the least upper bound on types.. We will use the notation $t_1 \uparrow t_2$ to mean the largest of $t_1$ and $t_2$. The rules are seen in Table 2, from which it can be seen that:

- The least upper bound of addition and multiplication is the least upper bound of each component.
- The least upper bound of numbers is the largest number.
- The least upper bound of variables is the maximum of the variables.
- The least upper bound of the maximum of some types, is just the maximum of all the types.
- The least upper bound of record-types is the record where each identifier has the least upper bound of the types in each record.
- For arrow-types, we assume that components will always have the same number and type of arguments. However, the output might change, as one component might take much longer time than another one with the same signature.

| Name | Partial order |
|---|---|
| (Composition) | $$\frac{\sqcup\{t_1,...,t_n\}=t \quad \sqcup\{t_{1'},...,t_{n'}\}=t'}{\sqcup\{t_1 \oplus t_{1'},...,t_n \oplus t_{n'}\}=t \oplus t'} \oplus \in \{+,\cdot\}$$ |
| (Number) | $$\frac{t=\max(n,...,m)}{\sqcup\{n,...,m\}=t} \; n,...,m \in \mathbb{N}$$ |
| (Variable) | $$\overline{\sqcup\{x,...,y\}=x \uparrow ... \uparrow y}$$ |
| (Number) | $$\frac{t=\max(n,...,m)}{\sqcup\{n,...,m\}=t} \; n,...,m \in \mathbb{N}$$ |
| (Max) | $$\overline{\sqcup\{t_1 \uparrow t'_1,...,t_n \uparrow t'_n\}=t_1 \uparrow t'_1 \uparrow ... \uparrow t_n \uparrow t'_n}$$ |
| (Record) | $$\frac{\text{for each } i \sqcup\{t_i,...,t'_i\}=t''_i}{\sqcup\left\{\{id_i:t_i\}_{i\in 1...n},...,\{id_i:t'_i\}_{i\in 1...n}\right\}=\{id_i:t''_i\}_{i\in 1...n}}$$ |
| (Arrow) | $$\frac{\sqcup\left\{t_{n_1},...,t_{n_m}\right\}=t'}{\sqcup\left\{t_1 \to ... \to t_{n-1} \to t_{n_1},...t_1 \to ... \to t_{(n-1)} \to t_{n_m}\right\}=t_1 \to ...t_{n-1} \to t'}$$ |

Table 2: The rules of the partial order of types

We will always assume that the equivalence rules are used until the minimum least upper bound is found. An example of this is if we have $\sqcup\{1+3, 3+1\}$. If we just applied the least upper bound, we would get that it was $3+3$. However, because we assume that we will use the equivalence rules to minimize the least upper bound, we use that: $1+3 \equiv 3+1$, and we therefore instead have: $\sqcup\{3+1, 3+1\} = 3+1$.

Like the partial order on types, the partial order on type environments is defined through the definition of the least upper bound on type environments. The partial order on environments is:

$$\sqcup(\{\Gamma_1,...,\Gamma_n\}) = \Gamma'$$

where

$$\Gamma'(id) = \sqcup\{t \mid \exists \Gamma'' \in \{\Gamma_1,...,\Gamma_n\}.t = \Gamma''(id)\}$$

This is a point-wise upper bound which means that the least upper bound takes the largest element possible for each *id*, from each of the environments. We also define a least element, 0, for which we have that:

$$0 \sqsubseteq t$$

This means that all types are greater than or equal to 0. This will be be used mostly for the soundness proof of the type system, to make sense of operations that in our abstract machine have no cost, but instead creates a construct that we need to type differently than a number in our type system.

## 3.3 Type Rules for Expressions

### 3.3.1 BinOp
The rule for binary operators takes both the cost of the two expressions into account, and also the cost of applying the binary operator. We only allow binary operations on simple types, and not arrow-types or record-types.

$$(\text{T-BinOp}) \frac{\Gamma_1 \vdash e : t_1 \quad \Gamma_1 \vdash e : t_2 \quad t_3 = bc(\mathsf{BinOp})}{\Gamma \vdash \mathsf{BinOp}\ e_1 e_2 : t_1 + t_2 + t_3} \ t_1, t_2 \notin \{t \to t, \{id : t\}\}$$

### 3.3.2 Component Definition
To type a component definition, which acts as a function, we introduce type variables $t_1...t_n$. If given these type-variables, the statements can be typed as $t_r$, we then get the arrow-type $t_1 \to ... \to t_n \to t_r$.

$$(\text{T-ComDef}) \frac{\Gamma[id_1 : t_1, ..., id_n : t_n] \vdash \vec{S} : t_r}{\Gamma \vdash \mathsf{CompDef}\ \vec{id}\ \vec{S} : t_1 \to ... \to t_n \to t_r}$$

The rest of the type rules for expressions can be found in Appendix C.1

## 3.4 Type Rules for Statements

### 3.4.1 While
The while-loop may be non-terminating, so its total cost may not be defined. However, we only account for terminating programs, so to counter this, we introduce a variable, $x$. This variable can be set to any value, but the best estimate would be the exact number of iterations of the loop.

Using this variable, $x$, we can calculate the cost of a while-loop as the product of $x$ and the sum of the cost of the conditional expression, $t_1$, and the loop body, $t_2$. Finally, to terminate the while-loop the expression is checked an additional time, which explains why the cost of the expression is added once more, as seen in (T-While).

$$(\text{T-While}) \frac{\Gamma_1 \vdash e : t_1 \quad \Gamma_1 \vdash \vec{S} : t_2 \triangleright \Gamma_2}{\Gamma_1 \vdash \mathsf{While}\ e\ \vec{S} : x \cdot (t_1 + t_2) + t_1 \triangleright \sqcup \{\Gamma_1, \Gamma_2\}}$$

As is seen, the environment is also updated to be the least upper bound of the environment before and after execution of the loop body ($\vec{S}$). This is necessary, as we do not know if the loop body is executed at all, so we must consider both cases in the resulting environment.

The while-loop is the only construct that introduces free variables, which means that the costs of the remaining rules can be estimated without assuming values to variables. It is not possible to avoid the complexity added by the while-loop, as it is required to ensure Turing-completeness. However, if the for-loop are used instead, variables in the type can be avoided.

### 3.4.2 For
In this subset of the language, we include a very simple version of the for-loop, where for each iteration, a variable, $id$, is assigned to a value exactly one larger than in the previous iteration. The first value of $id$ is $n_1$, and the last value of $id$, with which the loop runs, is $n_2$. The loop body, $\vec{S}$, is executed once for each value of $id$. The total cost of the execution of the loop body and the assignments to $id$ is shown in (T-FOR).

$$(\text{T-FOR}) \frac{\Gamma_1 \vdash \vec{S} : t_1 \triangleright \Gamma_2 \qquad n_1 \leq n_2}{\Gamma_1 \vdash \text{For } id \ n_1 \ n_2 \ \vec{S} : t_2 \cdot t_1 + t_2 \triangleright \Gamma_2} \ t_2 = (n_2 - n_1 + 1)$$

In (T-FOR), $t_2$ is the number of iterations of the loop, which is also the number of assignments to the variable $id$. Therefore, the total cost can be calculated as the the product of the number of iterations and the cost of the loop body, i.e., $t_2 \cdot t_1$ to which we must add the cost of the $t_2$ assignments to $id$. The cost of the assignments is simply $t_2$, which can be concluded from the rules (T-AssignSimple) and (T-Num).

### 3.4.3 If

When estimating the cost of the if-statement, we must handle both parts of the statement: the conditional expression, $e$, and the two branches, $\vec{S}_1$ and $\vec{S}_2$, out of which only one is executed. As the conditional expression is always evaluated, we add its cost. Regarding the two statements, we do not know which one is executed. Therefore, we choose to add the cost of the most expensive branch to the cost of the if-statement. This is formalized in (T-IF).

$$(\text{T-IF}) \frac{\begin{array}{c} \Gamma \vdash e : t_1 \\ \Gamma \vdash \vec{S}_1 : t_2 \triangleright \Gamma_1 \\ \Gamma \vdash \vec{S}_2 : t_3 \triangleright \Gamma_2 \end{array}}{\Gamma \vdash \text{If } e \ \vec{S}_1 \ \vec{S}_2 : t_1 + (t_2 \uparrow t_3) \triangleright \sqcup \{\Gamma_1, \Gamma_2\}}$$

As can be seen, the resulting environment is also updated to be the least upper bound of the two environments, $\Gamma_1$ and $\Gamma_2$, resulting from typing the two branches to preserve potential updates of the types in $\Gamma$.

## 3.5 Type Rule for Import

The import construct will have a cost of 2 plus the type of the imported file, as well as the cost of assigning each variable that you want to import from the file.

$$(\text{T-IMPORTSELECTED}) \frac{[] \vdash fg(\text{File}) : t \triangleright \Gamma'\left[\varepsilon' \to \{id_i : t_i\}_{i \in \{1..n..m\}}\right]}{\Gamma \vdash \text{Import } \vec{id}^n \ \text{File} : t + n + 2 \triangleright \Gamma[id_1 \to t_1, ..., id_n \to t_n, \varepsilon \to \{\}]}$$

The rest of the type rules for statements can be found in Appendix C.2

# 4 Soundness of the Type System

## 4.1 Soundness Theorem

To define soundness for our type system, we will first define the accumulated cost of executing a statement on the abstract machine.

**Definition 4** (Total cost): $S \stackrel{c}{\Longrightarrow} S'$ iff $S \stackrel{c_1}{\Longrightarrow} ... \stackrel{c_n}{\Longrightarrow} S'$ where $c = \sum_{i=1}^{n} c_i$

We need Definition 4 to express that a series of transitions from a statement $S$ to $S'$ has a cost, $c$, without explicitly stating all possible transitions between the two statements.

**Theorem 1** (Soundness): Given a file-getter $fg$ and a file $f$ if $\Gamma \vdash fg(f) : t(\vec{x})$ and a execution of the abstract machine $\langle fg(f) : nil, fg, [], [], nil, f : nil \rangle \overset{c}{\Longrightarrow} \langle nil, fg, [], [], nil, nil \rangle$ then $\exists \sigma \in \vec{x} \to \mathbb{N}$ such that $c \sqsubseteq \sigma(t(\vec{x}))$

Soundness in our type-system states that the type cost is greater than or equal to the cost of an execution of the program on the abstract machine. The problem here is that there exists a set of free variables $\vec{x}$ in the type $t$, where $\exists \sigma \in \vec{x} \to \mathbb{N}$ which is an instantiation of the free variables. Finding the function $\sigma$ is incalculable in the case of free variables generated by while loops, as the number of iterations is only known at runtime.

In the next section, we will go through a subset of the proof of soundness for the type system on the height of the derivation tree. The full proof can be found in Appendix D.

## 4.2 Proof of Soundness of Statements

### 4.2.1 While

$$(\text{T-While}) \; \frac{\Gamma_1 \vdash e : t_1 \quad \Gamma_1 \vdash \vec{S} : t_2 \triangleright \Gamma_2}{\Gamma_1 \vdash \text{While } e \; \vec{S} : x \cdot (t_1 + t_2) + t_1 \triangleright \sqcup \{\Gamma_1, \Gamma_2\}}$$

We know from the inductive hypothesis that $e : t_1$ and $\vec{S} : t_2$ means that the trace of $e$ and $\vec{S}$ being evaluated will cost at most $t_1 + t_2$. With this, from the abstract machine, we have the transition sequence:

$$\langle \text{While } e \; \vec{S}, fg, ls, es, vs, ss \rangle \Longrightarrow \langle e :: \text{While}' \; e \; \vec{S} :: rest, fg, ls, es, vs, ss \rangle$$
$$\overset{c_1}{\Longrightarrow} \langle \text{While}' e \; \vec{S} :: rest, fg, ls, es, v :: vs, ss \rangle \text{ where } v \neq 0$$
$$\Longrightarrow \langle \vec{S} :: e :: \text{While}' e \; \vec{S} :: rest, fg, ls, es, vs, ss \rangle$$
$$\overset{c_2}{\Longrightarrow} \langle e :: \text{While}' e \; \vec{S} :: rest, fg, ls, es, vs, ss \rangle$$

where each iteration $i$ goes through the trace starting from the second transition.

When the condition costs $c_1$ such that $c_1 \sqsubseteq t_1$, and the statement costs $c_2$ such that $c_2 \sqsubseteq t_2$, and $i \leq x$, then it holds for all iterations that $i \cdot (c_1 + c_2) \sqsubseteq x \cdot (t_1 + t_2)$, where the while statement condition holds.

When the condition does not hold, only the condition is evaluated in the abstract machine, and we continue to the remaining program "*rest*".

$$\langle e :: \text{While}' \; e \; \vec{S} :: rest, fg, ls, es, vs, ss \rangle$$
$$\overset{c_1}{\Longrightarrow} \langle \text{While}' e \; \vec{S} :: rest, fg, ls, es, v :: vs, ss \rangle \text{ where } v = 0$$
$$\Longrightarrow \langle rest, fg, ls, es, vs, ss \rangle$$

Adding the final while statement condition check we get that $(i \cdot (c_1 + c_2) + c_1) \sqsubseteq (x \cdot (t_1 + t_2) + t_1)$ where $i \leq x$. Since we can assign the free variable $x$ to any value as long as it is equal or greater than $i$, the soundness theorem holds for the While statement. This also implies that the best estimate for $x$ is $i$.

### 4.2.2 For

$$(\text{T-For}) \frac{\Gamma_1 \vdash \vec{S} : t_1 \triangleright \Gamma_2 \quad n_1 \leq n_2}{\Gamma_1 \vdash \text{For } id \ n_1 \ n_2 \ \vec{S} : t_2 \cdot t_1 + t_2 \triangleright \Gamma_2} \ t_2 = (n_2 - n_1 + 1)$$

We know from the inductive hypothesis that the for-loop body has a cost of $\vec{S} : t_1$ and runs $t_2 = n_2 - n_1 + 1$ iterations as per the side condition. With this, from the abstract machine, we have the transition sequence:

$$\langle \text{For } id \ n_1 n_2 \ \vec{S} :: rest, fg \ ls, es, vs, ss \rangle$$
$$\implies \langle \text{Let } id \ n_1 :: \vec{S} :: ... :: \text{Let } id \ n_i :: \vec{S} :: ... :: \text{Let } id \ n_2 :: \vec{S} :: rest, fg, ls, es, vs, ss \rangle$$
$$\text{where } n_i = \mathbb{N}^{-1}(\mathbb{N}(n_1) + i) \text{ and } \mathbb{N}(n_1) \leq \mathbb{N}(n_2)$$

In the trace, the for-loop is unrolled and the current index in the loop is bound for each iteration starting at $n_1$ and ending at $n_2$ giving a total of $c_2 = n_2 - n_1 + 1$ iterations of the for loop. Therefore, it must hold that $c_2 \sqsubseteq t_2$.

The statement of the iteration has a cost of $c_1$ such that $c_1 \sqsubseteq t_1$. As the index is bound for each iteration and the cost of a let binding is 1, it must hold that an additional cost per iteration of $t_2$ must be added to account for the bindings. Therefore it must hold that $c_2 \cdot c_1 + c_2 \sqsubseteq t_2 \cdot t_1 + t_2$

### 4.2.3 Component Call

$$(\text{T-CompCall}) \frac{\begin{array}{c} \Gamma \vdash e_{2_1} : t_{11} \\ \vdots \\ \Gamma \vdash e_{2_n} : t_{nn} \\ \Gamma \vdash e_1 : t_1 \to ... \to t_n \to t_r \end{array} \quad \begin{cases} t_{ii} \sqsubseteq t_i \mid t_i \in \{t \to t, \{id:t\}\} \\ t_i \sqsubseteq 0 \mid \text{otherwise} \end{cases} \text{ where } i \in 1...n}{\Gamma \vdash \text{CompCall } e_1 \vec{e_2} : t_r + t_a + n \triangleright \Gamma} \ t_a = \sum_{1...n} \begin{cases} 0 \mid t_{ii} \in \{t \to t, \{id:t\}\} \\ t_{ii} \mid \text{otherwise} \end{cases}$$

From the inductive hypothesis, we know that the statements in the closure of $e_1$ costs at most $t_r$, where $t_1$ to $t_n$ are bound to their concrete type. With this, the abstract machine can execute the transition, binding all the CompCall arguments.

$$\langle \text{CompCall } e_1 \vec{e_2} :: rest, fg, ls, es, vs, ss \rangle$$
$$\implies \langle e_1 :: \text{CompCall}' \ \vec{e_2} :: rest, fg, ls, es, vs, ss \rangle$$
$$\implies \langle \text{CompCall}' \ \vec{e_2} :: rest, fg \ ls, es, \left(\vec{id}, \vec{S}, s\right) :: vs, ss \rangle$$
$$\implies \langle e_1 :: ... :: e_n :: \text{PushScope } s :: \text{Bind } id_n :: ... :: \text{Bind } id_1 :: \vec{S} :: \text{PopScope} :: rest, fg, ls, es, vs, ss \rangle$$
$$\stackrel{c_1}{\implies} \langle \vec{S} :: \text{PopScope} :: rest, fg, ls, es, vs, s :: ss \rangle$$

The cost of binding the arguments is the cost of evaluating the expressions and the cost of binding the arguments denoted as $c_1$. The number of Bind's is $n$, where $n$ is the number of arguments from the CompCall typing judgement. Therefore since $c_1$ is the cost of both the binding and cost of the argument expressions then $c_1 \sqsubseteq t_a + n$ where $t_a$ is the cost of evaluating the argument expressions in the CompCall typing judgement. With this, the abstract machine can execute the transition of the components body.

$$\langle \vec{S} :: \text{PopScope} :: rest, fg, ls, es, vs, s :: ss \rangle$$
$$\stackrel{c_2}{\Longrightarrow} \langle \text{PopScope} :: rest, fg, ls, es, vs, s :: ss \rangle$$
$$\Longrightarrow \langle rest, fg, ls, es, vs, ss \rangle$$

The cost of the trace evaluating $\vec{S}$ is $c_2$ such that:

$$c_2 \sqsubseteq t_r \text{ given } \begin{cases} t_{ii} \sqsubseteq t_i & | \ t_i \in \{t \to t, \{id:t\}\} \\ t_i \sqsubseteq 0 & | \ \text{otherwise} \end{cases} \text{ where } i \in 1...n$$

Where the side condition ensures that cost of evaluating the expression is only counted once by setting the argument cost to 0, given that it is not an Object or a CompDef. Therefore it holds that $c_1 + c_2 \sqsubseteq t_r + t_a + n$.

### 4.2.4 If

$$(\text{T-If}) \ \frac{\begin{array}{c} \Gamma \vdash e : t_1 \\ \Gamma \vdash \vec{S}_1 : t_2 \triangleright \Gamma_1 \\ \Gamma \vdash \vec{S}_2 : t_3 \triangleright \Gamma_2 \end{array}}{\Gamma \vdash \text{If } e \ \vec{S}_1 \ \vec{S}_2 : t_1 + (t_2 \uparrow t_3) \triangleright \sqcup \{\Gamma_1, \Gamma_2\}}$$

We know from the inductive hypothesis that $e : t_1$ and $\vec{S_1} : t_2$ and $\vec{S_2} : t_3$ means that the trace of $e$ and either $\vec{S_1}$ or $\vec{S_2}$ being evaluated will cost at most $t_1 + (t_2 \uparrow t_3)$. With this, from the abstract machine, we can go from:

$$\langle \text{If } e_1 \vec{S_1} \vec{S_2} :: rest, fg, ls, es, vs, ss \rangle \Longrightarrow \langle e_1 :: \text{Branch } \vec{S_1} \vec{S_2} :: rest, fg, ls, es, vs, ss \rangle \stackrel{c_1}{\Longrightarrow}$$
$$\langle \text{Branch } \vec{S_1} \vec{S_2} :: rest, fg, ls, es, v :: vs, ss \rangle$$

Where $c_1$ is the cost of evaluating the condition, such that $c_1 \sqsubseteq t_1$.

Depending on whether the condition holds, there are two traces.

$$\langle \text{Branch } \vec{S_1} \vec{S_2} :: rest, fg, ls, es, v :: vs, ss \rangle \Longrightarrow \langle \vec{S_1} :: rest, fg, ls, es, vs, ss \rangle \text{ where } v \neq 0$$

$$\langle \text{Branch } \vec{S_1} \vec{S_2} :: rest, fg, ls, es, v :: vs, ss \rangle \Longrightarrow \langle \vec{S_2} :: rest, fg, ls, es, vs, ss \rangle \text{ where } v = 0$$

The cost of the trace evaluating $\vec{S_1}$ will be $c_2$ and the trace for $\vec{S_2}$ will be $c_3$ such that $c_2 \sqsubseteq t_2$ and $c_3 \sqsubseteq t_3$. Since it is not known which branch is evaluated, the worst case is selected such that $c_2 \sqsubseteq (t_2 \uparrow t_3)$ and $c_3 \sqsubseteq (t_2 \uparrow t_3)$. Adding the cost of the condition of the if statement we get that $c_1 + c_2 \sqsubseteq t_1 + (t_2 \uparrow t_3)$ and $c_1 + c_3 \sqsubseteq t_1 + (t_2 \uparrow t_3)$ therefore the soundness theorem holds for both branches of the (T-If) statement rule.

## 4.3 Proof of Soundness of Imports

### 4.3.1 Import Selected

$$(\text{T-ImportSelected}) \ \frac{[] \vdash fg(\text{File}) : t \triangleright \Gamma'\left[\varepsilon' \to \{id_i : t_i\}_{i \in \{1..n..m\}}\right]}{\Gamma \vdash \text{Import } \vec{id}^n \text{ File} : t + n + 2 \triangleright \Gamma[id_1 \to t_1, ..., id_n \to t_n, \varepsilon \to \{\}]}$$

From the inductive hypothesis we know that $fg(\text{File})$ will cost at most $t$. If we look at the trace of the abstract machine, for (R-ImportSelected), we have:

$$\langle \vec{id} \ f :: rest, fg, ls, es, vs, ss \rangle \stackrel{2}{\Longrightarrow} \langle fg(f) :: \text{PopScope} :: \overline{\text{BindSelected } \vec{id}} :: \text{EmptyExports} :: \\ rest, fg, ls, es, vs, f :: ss \rangle$$

Giving us an initial cost of 2. We know that when $fg(f)$ is expanded, it results in a sequence of instructions. The execution of this transition sequence will have a cost of $c$ such that $c \sqsubseteq t$. Then, if we look at the next step of the trace, we have that pop-scope costs 0.

$$\langle \mathsf{PopScope} :: \overline{\mathsf{BindSelected}\ id} :: \mathsf{EmptyExports} :: rest, fg, ls, es, vs, f :: ss\rangle \Longrightarrow$$
$$\langle \overline{\mathsf{BindSelected}\ id} :: \mathsf{EmptyExports} :: rest, fg, ls, es, vs, ss\rangle$$

We can expand the trace into $\langle \mathsf{BindSelected}\ id :: ... :: \mathsf{BindSelected} :: \mathsf{EmptyExports} :: rest, fg, ls, es, vs, ss\rangle$ where the number of BindSelected corresponds to $m = |\overline{\mathsf{BindSelected}\ id}|$. The trace for a single BindSelected in the abstract machine is as follows:

$$\langle \mathsf{BindSelected}\ id :: rest, fg, ls, es, vs, s :: ss\rangle \overset{1}{\Longrightarrow} \langle rest, fg, ls[(s, id) \mapsto es(id)], es, vs, s :: ss\rangle$$

We observe that the cost of each of the $m$ BindSelected is 1, and as $m$ is equivalent to the number of $id$'s, it must hold that $m \sqsubseteq n$. We can observe that the final trace of the transition sequence until "$rest$" has a cost of 0.

$$\langle \mathsf{EmptyExports} :: rest, fg, ls, es, vs, ss\rangle \Longrightarrow \langle rest, fg, ls, [], vs, ss\rangle$$

Since $m \sqsubseteq n$ and $c \sqsubseteq t$, it must hold that $c + m \sqsubseteq t + n$. With the initial cost of 2 in the abstract machine, which can be observed in the type judgement for the (T-IMPORTSELECTED) rule, it must hold that $c + m + 2 \sqsubseteq t + n + 2$.

## 4.4 Proof of Soundness of Expressions

### 4.4.1 BinOp

$$(\text{T-BINOP}) \frac{\Gamma_1 \vdash e : t_1 \qquad \Gamma_1 \vdash e : t_2 \qquad t_3 = bc(\mathsf{BinOp})}{\Gamma \vdash \mathsf{BinOp}\ e_1 e_2 : t_1 + t_2 + t_3} \quad t_1, t_2 \notin \{t \to t, \{id : t\}\}$$

The BinOp rule is typed with the cost of $t_1 + t_2 + t_3$. From the inductive hypothesis, we know that $e_1$ can be reduced with a cost of at most $t_1$, and $e_2$ can be reduced with a cost of at most $t_2$. We also know that $t_3$ is equal to $bc(\mathsf{BinOp})$.

$$\langle \mathsf{BinOp}\ e_1\ e_2 :: rest, fg, ls, es, vs, ss\rangle \underset{c_2}{\Longrightarrow} \langle e_1 :: e_2 :: \mathsf{BinOp} :: rest, fg, ls, es, vs, ss\rangle \overset{c_1}{\underset{c_3}{\Longrightarrow}}$$
$$\langle e_2 :: \mathsf{BinOp} :: rest, fg, ls, es, v_1 :: vs, ss \Longrightarrow (\mathsf{BinOp} :: rest, fg, ls, es, v_2 :: v_1 :: vs, ss) \Longrightarrow$$
$$(rest, fg, ls, es, v_3 :: vs, ss) \text{ where } v_3 = \mathsf{BinOp}(v_1, v_2) \text{ and } c_3 = bc(\mathsf{BinOp})$$

The execution of the two sub expressions has a cost of $c_1 + c_2$ such that $c_1 + c_2 \sqsubseteq t_1 + t_2$ and the BinOp has a cost of $c_3$ such that $c_3 \sqsubseteq t_3$. Therefore it must hold that $c_1 + c_2 + c_3 \sqsubseteq t_1 + t_2 + t_3$.

### 4.4.2 Component Definition

$$(\text{T-COMDEF}) \frac{\Gamma[id_1 : t_1, ..., id_n : t_n] \vdash \vec{S} : t_r}{\Gamma \vdash \mathsf{CompDef}\ \vec{id}\ \vec{S} : t_1 \to ... \to t_n \to t_r}$$

The CompDef rule is typed as an arrow type $t_1 \to ... \to t_n \to t_r$ where $t_1$ to $t_n$ are free variables corresponding to the arguments to the component. When examining the abstract machine trace for CompDef, we observe:

$$\langle \mathsf{CompDef}\ \vec{id}\ \vec{S} :: rest, fg, ls, es, vs, s :: ss\rangle \Longrightarrow \langle rest, fg, ls, es, \left(\vec{id}, \vec{S}, s\right) :: vs, s :: ss\rangle$$

The cost of creating a component is $c = 0$, and therefore it must hold that $c \sqsubseteq t$.

# 5 Application of Type system

## 5.1 Abstract Machine
We have implemented our abstract machine in Haskell, for which the full source code is available at [11]. The functionality of the abstract machine is implemented in a function called run, of which a snippet can be seen on Listing 3. The function takes an abstract machine configuration as input and returns a state transformer. The value evaluated in our state is a map representing our locals, and the state is a tuple containing the cost and trace of the reduction. We use a state transformer on the `Either String` monad since the `run` function can fail, as it is possible to make a abstract machine configuration where no reduction rule can be applied.

The code snippet in Listing 3 shows the implementation of the (R-ImportSelected) reduction rule. First, we add to the state by pushing the current reduction rule used to the trace and adding the cost of this particular rule to the total cost. We can do a lookup for the source file in our file getter `fg` by using the helper function `fileLookup`. Lastly, we recursively call the `run` function with a modified abstract machine configuration using the same modification as seen in the (R-ImportSelected) reduction rule in Table 3.

```
run ::
    AMState ->
    StateT (Int, Trace) (Either String) (M.Map (String, String) RuntimeValue)
run st@(AMState (Import (AST.ImportList idents f) : stack) fg ls es vs ss) = do
    pushTrace st "R-ImportSelected"
    addCost (+2)
    src <- lift $ fileLookup f fg
    run (AMState (concat [ [SrcFile src]
                         , [PopScope]
                         , map BindExport idents
                         , [EmptyExports]
                         , stack
                         ] ) fg ls es vs (f : ss))
```

Listing 3: Snippet of `run` function in Haskell implementation of the abstract machine.

If we run the abstract machine on the example on Listing 2 (main.jsx), we get that the reduction has a cost of 12, with the trace seen on Figure 6 in Appendix E, and locals seen on Figure 5.

$$(/\text{main.jsx}, \text{func}) \to ((\text{prop}), [\textsf{Assign } y \text{ prop} + 2], /\text{main.jsx})$$
$$(/\text{main.jsx}, \text{prop}) \to 0$$
$$(/\text{main.jsx}, x) \to 0$$
$$(/\text{main.jsx}, y) \to 2$$
$$(/\text{simpleWhile.jsx}, x) \to 0$$

Figure 5: Locals for the example on Listing 2

Figure 5 is of the form $ls[(s, id) \mapsto v]$, where the scope is an identifier of the scope it is bound in, the $id$ is the identifier we bind to, and $v$ is the value of said identifier at the end of the program.

### 5.2 Type Inference

We will now outline a way to use the type system to infer a type for a given file. The problem can be defined as:

Given a statement $S$, and a type $t$, can we find an instantiation of the free type-variables in $t$, so that $S$ is well-typed.

The most interesting part of the inference is the arrow type. Here, we need to gather the constraints for the input type, as we can only infer that type from all the places the component is called. This means that to gain the type for a file, we need to gather all the constraints, of the form $t_a \sqsubseteq t_b$, and solve for the smallest possible types.

An example could be, if we have a component definition, we can only initially say that its type is $t_1 \rightarrow t_2$. Then if one place in the program, we call the component with another component of type $(0 \rightarrow 1)$, then we get the constraint $(0 \rightarrow 1) \sqsubseteq t_1$. If we somewhere else call the component with an expression with type $(0 \rightarrow 2)$, then we also get the constraint $(0 \rightarrow 2) \sqsubseteq t_1$. To solve these constraints, we take the least upper bound of $(0 \rightarrow 1)$ and $(0 \rightarrow 2)$, which will be $(0 \rightarrow 2)$. Once we have this, we can then finally type the original component, since we now can type the type variable $t_1$ as $0 \rightarrow 2$. A strategy for gathering the constraints can be found from the soundness proof, as it shows a way to get an over-approximation of the cost. Essentially, for each type rule, you need to gather the constraints for each sub-expression and each sub-statement, to gain the constraint of the given construct.

### 5.3 Implementation of Constraint Gathering

We have also implemented this constraint gathering in Haskell, which brings us ones step closer to inferring the type of a file. Formally this gathering can be thought of as a transition system formulated as a syntax directed set of inference rule of the form $\Gamma \vdash \text{Stm} \rightarrow \mathcal{C} \rhd \Gamma$, where $\mathcal{C}$ is a set of constraints. The transition system works as follows: On every statement, except for a component call statement, we take the union of all the constraints found in the sub-statements and sub-expressions. For let and assign statements we type check the sub-expression and add a binding from the variable to the type in $\Gamma$. If the expression is a component, we use type variables, within the arrow type. Lastly for a component call, we type the component-expression and the argument-expressions. We can then add the constraints seen in our typing rule for component calls (T-CompCall). As an an example, one could define the inference rule for gathering constraints in a component call as seen below. It should be noted that we use $x$ here to denote a type variable, which means that during type checking this will not be a concrete type, but rather a type containing type variables that we are trying to infer.

$$\frac{\Gamma \vdash e_1 : x_1 \rightarrow ... \rightarrow x_n \rightarrow x_r \quad \Gamma \vdash e_{2_1} : t_1 \ ... \ \Gamma \vdash e_{2_n} : t_n}{\Gamma \vdash \textsf{CompCall } e_1 \vec{e_2} \rightarrow \{t_1 \sqsubseteq x_1, ..., t_n \sqsubseteq x_n\} \rhd \Gamma}$$

The implementation follows this interpretation very closely, as can be seen on Listing 4, which shows how constraint gathering is implemented for a component call. If we run the constraint gathering on the example in Listing 2, we see that only a single constraint is generated namely $((0+0)+0) \sqsubseteq \text{prop}$. For this example, the constraints gathered are rather easy to solve. However, one could imagine an example that makes use of nested component calls, components with multiple arguments and calls to the same component with arguments of varying types, which all will contribute to the complexity of the constraints.

```
1   stmtConstraints (CompCall e es) = do
2       env <- get
3       t <- lift $ evalStateT (exprTypeCheck e (E_Type <$> env)) 0
4       ts <- lift $ evalStateT (exprsTypeCheck es (E_Type <$> env)) 0
5       case t of
6           Function params _ -> do
7               lift $ sat
8                   (length params == length ts)
9                   "stmtConstraints: compCall: wrong number of arguments"
10              let cs = zipWith Lub (map show ts) params
11              return cs
12          _ -> lift $ Left "stmtConstraints: compCall: not a function"
```

Listing 4: Snippet of the constraint gathering implementation, showing how constraints are generated for a component call.

# 6 Discussion

## 6.1 Results

As shown, the type system will over-approximate the cost of executing the abstract machine on a file, given a variable assignment that is larger than or equal to the number of times the while loops are executed. With the proof of soundness, we have shown that the type system is sound, meaning that the cost of all well-typed programs will be over-approximated. Furthermore, if the file does not include while loops, but only use for loops, which has an explicit number of iterations, the cost estimate will not need to introduce variables, and therefore, we can find an over-approximation without unknowns. We have also implemented the abstract machine and a constraint-gatherer, to infer the type of a program.

Since JavaScript XML is primarily used for internet and web-development, use cases for this type system could be to preemptively infer the cost of executing files, so that you could begin executing files with a high cost, before they are needed. This could potentially make a user interface more responsive, as perhaps the change of a button press could already begin to be calculated, when the user is hovering over the button, but before they press it. Pre-loading is used in other places, such as browsers loading the webpage when the cursor is close to its hyperlink. It could also be used to find the more costly parts of a program, indicating which part of a code base that have the most potential to be improved upon. It can also give an estimate on a complexity of the algorithms in a file, as the type might include variables, that could indicate what degree of a polynomial the complexity of a file is. The complexity would not be sound, as we have no way for example to create exponential complexity with our type system.

## 6.2 Future Work

A simple extension of our current abstract machine is to include scopes in loops and conditionals. We deemed this unnecessary to show the theoretical capabilities of the type system.

A larger extension of our work, would be to expand the smaller language, so that it becomes the full JavaScript XML language. This means that our type system would be able to be implemented and actually cost-approximate all of a JavaScript XML file. However, this is outside the scope of this project.

Another way to extend this work would be to include a file-environment. This is because some files might already be loaded, and their exported values are already found. This means that we do not need to execute the imported file, as we already have the result of executing it. Therefore, a file that imports already loaded files should have its cost set to a much lower value, when over-approximating the cost of executing a file. This would be a rather simple extension, but we deemed it outside the scope of this project.

# A Abstract Machine

Rewriting the example from Section 1 using our abstract syntax for JavaScript XML, we get the following.

```
1  Let (x) (3)
2  while (x) (Assign (x) (Sub (x) (1)))
3  x //export
```

Listing 5: Rewriting /simpleWhile.jsx

```
1  x /simpleWhile.jsx //import
2  Let (y) (0)
3  Let (func) (CompDef (prop) (Assign (y) (Sub (prop) (2)))
4  CompCall (func) (x)
```

Listing 6: Rewriting /main.jsx

As we can not return values from functions, we have to create the variable y first, and then assign it inside of the function func, if we want to update the value of y with the function call in Listing 6 line 4. As we will see in the Section 2.4, y could (currently) also have been introduced in the while loop, given our scope rules for (R-While).

# B Reduction Rules

The rule (R-Num) reduces a number literal on the instruction stack by pushing the value $v$ of the syntactic element $n$ on to the value stack. Similarly the rule (R-Var) reduces a variable reference $id$ by looking up its value in locals, and then pushing the value on to the value stack.

(R-Proj) reduces a projection by first looking up the value of $id_1$ in locals, if that value is an object we lookup the value of the field $id_2$ and push the field' value on to the value stack.

The (R-ImportAll) rule handles the import of every exported variable from a file. This is decomposed into first evaluating the file under a new scope, then popping the scope, binding all the exports to some identifier and finally emptying the exports.

The rules (R-BindAll) and (R-BindSelected) concerns themselves with mapping to locals. (R-BindAll) binds a mapping to the locals, from an identifier and the current scope to the current exports. (R-BindSelected) also adds a new mapping to the locals from an identifier and the current scope, but the it maps to the value of an exported variable with the same identifier.

(R-Export) adds a mapping to the exports from an identifier to the value that the identifier has in the locals under the current scope. (R-EmptyExports) replaces the current exports with an empty function.

The (R-CompCall) rules handle a component call by first decomposing the call into evaluating the expression $e_1$ for the component closure, and then applying the arguments $\vec{e_2}$. When a CompCall' construct is reached the component closure must be on the value stack, we can then add a Let construct for each identifier argument pair, we can now execute the body of the component $\vec{S}$ and we add a PopScope construct to the instruction stack, to exit out of the function.

| Name | Rule | Side condition |
|------|------|----------------|
| (R-SrcFile) | $\langle \overrightarrow{\text{Import}}\ \vec{S}\ \overrightarrow{\text{Export}} :: rest, fg, ls, es, vs, ss \rangle$ | |
| | $\overset{0}{\Longrightarrow} \langle \mathsf{Import}_1 :: ... :: \mathsf{Import}_n :: S_1 :: ... ::$ $S_m :: \mathsf{Export}_1 :: ... :: \mathsf{Export}_j ::$ $rest, fg, ls, es, vs, ss \rangle$ | |
| (R-Num) | $\langle n :: rest, fg, ls, es, vs, ss \rangle$ $\overset{0}{\Longrightarrow} \langle rest, fg, ls, es, v :: vs, ss \rangle$ | $v = \mathbb{N}(n)$ |
| (R-BinOp1) | $\langle \mathsf{BinOp}\ e_1\ e_2 :: rest, fg, ls, es, vs, ss \rangle$ $\overset{0}{\Longrightarrow} \langle e_1 :: e_2 :: \mathsf{BinOp} :: rest, fg, ls, es, vs, ss \rangle$ | |
| (R-BinOp2) | $\langle \mathsf{BinOp} :: rest, fg, ls, es, v_2 :: v_1 :: vs, ss \rangle$ $\overset{c}{\Longrightarrow} \langle rest, fg, ls, es, v_3 :: vs, ss \rangle$ | $v_3 = \mathsf{BinOp}\ (v_1, v_2)$ $c = bc\ (\mathsf{BinOp})$ |
| (R-Var) | $\langle \mathsf{Var}\ id :: rest, fg, ls, es, vs, s :: ss \rangle$ $\overset{0}{\Longrightarrow} \langle rest, fg, ls, es, ls(s, id) :: vs, s :: ss \rangle$ | |
| (R-Proj) | $\langle \mathsf{Proj}\ id_1 id_2 :: rest, fg, ls, es, vs, s :: ss \rangle$ $\overset{0}{\Longrightarrow} \langle rest, fg, ls, es, \mathsf{Obj}(id_2) :: vs, s :: ss \rangle$ | $\mathsf{Obj} = ls(s, id_1)$ $\mathsf{Obj} :: (id \mapsto v)$ |
| (R-CompDef) | $\langle \mathsf{CompDef}\ \vec{id}\ \vec{S} :: rest, fg, ls, es, vs, s :: ss \rangle$ $\overset{0}{\Longrightarrow} \langle rest, fg, ls, es, (\vec{id}, \vec{S}, s) :: vs, s :: ss \rangle$ | |
| (R-If) | $\langle \mathsf{If}\ e_1 \overrightarrow{S_1} \overrightarrow{S_2} :: rest, fg, ls, es, vs, ss \rangle$ $\overset{0}{\Longrightarrow} \langle e_1 :: \mathsf{Branch}\ \overrightarrow{S_1}\ \overrightarrow{S_2} ::$ $rest, fg, ls, es, vs, ss \rangle$ | |
| (R-IfTrue) | $\langle \mathsf{Branch}\ \overrightarrow{S_1}\ \overrightarrow{S_2} :: rest, fg, ls, es, v :: vs, ss \rangle$ $\overset{0}{\Longrightarrow} \langle \vec{S}_1 :: rest, fg, ls, es, vs, ss \rangle$ | $v \neq 0$ |
| (R-IfFalse) | $\langle \mathsf{Branch}\ \overrightarrow{S_1}\ \overrightarrow{S_2} :: rest, fg, ls, es, v :: vs, ss \rangle$ $\overset{0}{\Longrightarrow} \langle \vec{S}_2 :: rest, fg, ls, es, vs, ss \rangle$ | $v = 0$ |
| (R-While) | $\langle \mathsf{While}\ e\ \vec{S}, fg, ls, es, vs, ss \rangle$ $\overset{0}{\Longrightarrow} \langle e :: \mathsf{While}'\ e\ \vec{S} :: rest, fg, ls, es, vs, ss \rangle$ | |
| (R-WhileTrue) | $\langle \mathsf{While}'\ e\ \vec{S} :: rest, fg, ls, es, v :: vs, ss \rangle$ $\overset{0}{\Longrightarrow} \langle \vec{S} :: e :: \mathsf{While}'\ e\ \vec{S} ::$ $rest, fg, ls, es, vs, ss \rangle$ | $v \neq 0$ |
| (R-WhileFalse) | $\langle \mathsf{While}'\ e\ \vec{S} :: rest, fg, ls, es, v :: vs, ss \rangle$ $\overset{0}{\Longrightarrow} \langle rest, fg, ls, es, vs, ss \rangle$ | $v = 0$ |
| (R-Let) | $\langle \mathsf{Let}\ id\ e :: rest, fg, ls, es, vs, ss \rangle$ $\overset{0}{\Longrightarrow} \langle e :: \mathsf{Bind}\ id :: rest, fg, ls, es, vs, ss \rangle$ | |
| (R-Assign) | $\langle \mathsf{Assign}\ id\ e :: rest, fg, ls, es, vs, ss \rangle$ $\overset{0}{\Longrightarrow} \langle e :: \mathsf{Bind}\ id :: rest, fg, ls, es, vs, ss \rangle$ | |
| (R-Bind) | $\langle \mathsf{Bind}\ id :: rest, fg, ls, es, v :: vs, s :: ss \rangle$ $\overset{1}{\Longrightarrow} \langle rest, fg, ls[(s, id) \mapsto v], es, vs, s :: ss \rangle$ | |
| (R-ImportAll) | $\langle id\ f :: rest, fg, ls, es, vs, ss \rangle$ | |
| | $\overset{2}{\Longrightarrow} \langle fg(f) :: \mathsf{PopScope} :: \mathsf{BindAll}\ id ::$ $\mathsf{EmptyExports} :: rest, fg, ls, es, vs, f :: ss \rangle$ | |
| (R-ImportSelected) | $\langle \vec{id}\ f :: rest, fg, ls, es, vs, ss \rangle$ $\overset{2}{\Longrightarrow} \langle fg(f) :: \mathsf{PopScope} :: \mathsf{BindSelected}\ id_1 ::$ $... :: \mathsf{BindSelected}\ id_n :: \mathsf{EmptyExports} ::$ $rest, fg, ls, es, f :: ss \rangle$ | |
| (R-PopScope) | $\langle \mathsf{PopScope} :: rest, fg, ls, es, vs, s :: ss \rangle$ $\overset{0}{\Longrightarrow} \langle rest, fg, ls, es, vs, ss \rangle$ | |
| (R-PushScope) | $\langle \mathsf{PushScope}\ s :: rest, fg, ls, es, vs, ss \rangle$ $\overset{0}{\Longrightarrow} \langle rest, fg, ls, es, vs, s :: ss \rangle$ | |
| (R-EmptyExports) | $\langle \mathsf{EmptyExports} :: rest, fg, ls, es, vs, ss \rangle$ $\overset{0}{\Longrightarrow} \langle rest, fg, ls, [], vs, ss \rangle$ | |
| (R-BindAll) | $\langle \mathsf{BindAll}\ id :: rest, fg, ls, es, vs, s :: ss \rangle$ $\overset{1}{\Longrightarrow} \langle rest, fg, ls[(s, id) \mapsto es], es, vs, s :: ss \rangle$ | |
| (R-BindSelected) | $\langle \mathsf{BindSelected}\ id :: rest, fg, ls, es, vs, s :: ss \rangle$ $\overset{1}{\Longrightarrow} \langle rest, fg, ls[(s, id) \mapsto es(id)], es, vs, s ::$ $ss \rangle$ | |
| (R-Export) | $\langle \mathsf{Export}\ id :: rest, fg, ls, es, vs, s :: ss \rangle$ $\overset{1}{\Longrightarrow} \langle rest, fg, ls, es[id \mapsto ls(s, id)], vs, s :: ss \rangle$ | |
| (R-CompCall) | $\langle \mathsf{CompCall}\ e_1 \vec{e}_2 :: rest, fg, ls, es, vs, ss \rangle$ $\overset{0}{\Longrightarrow} \langle e_1 :: \mathsf{CompCall'}\ \vec{e}_2 ::$ $rest, fg, ls, es, vs, ss \rangle$ | |
| (R-CompCallPrime) | $\langle \mathsf{CompCall'}\ \vec{e} :: rest, fg\ ls, es, (\vec{id}, \vec{S}, s) ::$ $vs, ss \rangle$ $\overset{0}{\Longrightarrow} \langle e_1 :: ... :: e_n :: \mathsf{PushScope}\ s ::$ $\mathsf{Bind}\ id_n :: ... :: \mathsf{Bind}\ id_1 :: S_1 :: ... :: S_m ::$ $\mathsf{PopScope} :: rest, fg, ls, es, vs, ss \rangle$ | |
| (R-For) | $\langle \mathsf{For}\ id\ n_1 n_2\ \vec{S} :: rest, fg\ ls, es, vs, ss \rangle$ $\overset{0}{\Longrightarrow} \langle \mathsf{Let}\ id\ n_1 :: S :: ... :: \mathsf{Let}\ id\ n_i :: S ::$ $... :: \mathsf{Let}\ id\ n_2 :: S :: rest, fg, ls, es, vs, ss \rangle$ | $n_i = \mathbb{N}^{-1}(\mathbb{N}(n_1) + i)$ $\mathbb{N}(n_1) \leq \mathbb{N}(n_2)$ |

Table 3: Reduction rules

# C Type Rules

## C.1 Expressions

$$(\text{T-Num}) \frac{}{\Gamma \vdash n : 0}$$

$$(\text{T-Var}) \frac{}{\Gamma \vdash id : t}\ t = \Gamma(id)$$

$$(\text{T-BinOp}) \frac{\Gamma_1 \vdash e : t_1 \quad \Gamma_1 \vdash e : t_2 \quad t_3 = bc(\text{BinOp})}{\Gamma \vdash \text{BinOp } e_1 e_2 : t_1 + t_2 + t_3} t_1, t_2 \notin \{t \to t, \{id : t\}\}$$

$$(\text{T-Proj}) \frac{\Gamma(id_1) = \{l_i : t_i\}_{i \in I}}{\Gamma \vdash \text{Proj } id_1 id_j : t_j} \text{ where } j \in I$$

$$(\text{T-ComDef}) \frac{\Gamma[id_1 : t_1, ..., id_n : t_n] \vdash \vec{S} : t_r}{\Gamma \vdash \text{CompDef } \vec{id} \; \vec{S} : t_1 \to ... \to t_n \to t_r}$$

## C.2 Statements

$$(\text{T-Comp}) \frac{\Gamma_0 \vdash S_1 : t_1 \triangleright \Gamma_1 ... \Gamma_{n-1} \vdash S_n : t_n \triangleright \Gamma_n}{\Gamma_0 \vdash \vec{S} : t_1 + ... + t_n \triangleright \Gamma_n}$$

$$(\text{T-While}) \frac{\Gamma_1 \vdash e : t_1 \quad \Gamma_1 \vdash \vec{S} : t_2 \triangleright \Gamma_2}{\Gamma_1 \vdash \text{While } e \; \vec{S} : x \cdot (t_1 + t_2) + t_1 \triangleright \sqcup \{\Gamma_1, \Gamma_2\}}$$

$$(\text{T-For}) \frac{\Gamma_1 \vdash \vec{S} : t_1 \triangleright \Gamma_2 \quad n_1 \leq n_2}{\Gamma_1 \vdash \text{For } id \; n_1 \; n_2 \; \vec{S} : t_2 \cdot t_1 + t_2 \triangleright \Gamma_2} t_2 = (n_2 - n_1 + 1)$$

$$(\text{T-CompCall}) \frac{\begin{array}{c}\Gamma \vdash e_{2_1} : t_{11} \\ \vdots \\ \Gamma \vdash e_{2_n} : t_{nn} \\ \Gamma \vdash e_1 : t_1 \to ... \to t_n \to t_r\end{array} \quad \begin{cases} t_{ii} \sqsubseteq t_i \mid t_i \in \{t \to t, \{id:t\}\} \\ t_i \sqsubseteq 0 \mid \text{otherwise}\end{cases} \text{ where } i \in 1...n}{\Gamma \vdash \text{CompCall } e_1 \vec{e_2} : t_r + t_a + n \triangleright \Gamma} t_a = \sum_{1...n} \begin{cases} 0 \mid t_{ii} \in \{t \to t, \{id:t\}\} \\ t_{ii} \mid \text{otherwise}\end{cases}$$

$$(\text{T-If}) \frac{\begin{array}{c}\Gamma \vdash e : t_1 \\ \Gamma \vdash \vec{S_1} : t_2 \triangleright \Gamma_1 \\ \Gamma \vdash \vec{S_2} : t_3 \triangleright \Gamma_2\end{array}}{\Gamma \vdash \text{If } e \; \vec{S_1} \; \vec{S_2} : t_1 + (t_2 \uparrow t_3) \triangleright \sqcup \{\Gamma_1, \Gamma_2\}}$$

$$(\text{T-LetSimple}) \frac{\Gamma \vdash e : t_1}{\Gamma \vdash \text{Let } id \; e : 1 + t_1 \triangleright \Gamma[id \to 0]} t_1 \notin \{t_1 \to t_2, \{id : t_3\}\}$$

$$(\text{T-LetAdvanced}) \frac{\Gamma \vdash e : t_1}{\Gamma \vdash \text{Let } id\ e : 1 \triangleright \Gamma[id \to t_1]} t_1 \in \{t_1 \to t_2, \{id : t_3\}\}$$

$$(\text{T-AssignSimple}) \frac{\Gamma \vdash e : t_1 \qquad id \in \text{dom}(\Gamma)}{\Gamma \vdash \text{Assign } id\ e : 1 + t_1 \triangleright \Gamma[id \to 0]} t_1 \notin \{t_1 \to t_2, \{id : t_3\}\}$$

$$(\text{T-AssignAdvanced}) \frac{\Gamma \vdash e : t_1 \qquad id \in \text{dom}(\Gamma)}{\Gamma \vdash \text{Assign } id\ e : 1 \triangleright \Gamma[id \to t_1]} t_1 \in \{t_1 \to t_2, \{id : t_3\}\}$$

$$(\text{T-SrcFile}) \frac{\begin{array}{c}\Gamma_0 \vdash im_1 : t_{1_1} \triangleright \Gamma^1 ... \Gamma^{n-1} \vdash im_n : t_{1_n} \triangleright \Gamma^n \\ \Gamma^n \vdash \vec{S} : t_{2_1} + ... + t_{2_m} \triangleright \Gamma^{n+m} \\ \Gamma^{n+m} \vdash ex_1 : t_{3_1} \triangleright \Gamma^{n+m+1} ... \Gamma^{n+m+l-1} \vdash ex_l : t_{3_l} \triangleright \Gamma^{n+m+l} \end{array}}{\Gamma_0 \vdash \text{SrcFile } \vec{im}\ \vec{S}\ \vec{ex} : t_{1_1} + ... + t_{1_n} + t_{2_1} + ... + t_{2_m} + t_{3_1} + ... + t_{3_l} \triangleright \Gamma^{n+m+l}}$$

## C.3 Imports

$$(\text{T-ImportAll}) \frac{[] \vdash fg(\text{File}) : t \triangleright \Gamma'}{\Gamma \vdash \text{Import } id\ \text{File} : t + 3 \triangleright \Gamma[id \to \Gamma'(\varepsilon), \varepsilon \to \{\}]}$$

$$(\text{T-ImportSelected}) \frac{[] \vdash fg(\text{File}) : t \triangleright \Gamma'\left[\varepsilon' \to \{id_i : t_i\}_{i \in \{1..n..m\}}\right]}{\Gamma \vdash \text{Import } \vec{id}^n\ \text{File} : t + n + 2 \triangleright \Gamma[id_1 \to t_1, ..., id_n \to t_n, \varepsilon \to \{\}]}$$

## C.4 Export

$$(\text{T-Export}) \frac{t = \Gamma(id) \qquad \{id_i : t_i\}_{i \in \{1..n\}} = \Gamma(\varepsilon)}{\Gamma \vdash \text{Export } id : 1 \triangleright \Gamma[\varepsilon \to ex] \quad ex = \{id_1 : t_1, ..., id_n : t_n, id : t\}_{i \in 1...n+1}}$$

# D Soundness Proof

**Definition 4** (Total cost): $S \stackrel{c}{\Longrightarrow} S'$ iff $S \stackrel{c_1}{\Longrightarrow} ... \stackrel{c_n}{\Longrightarrow} S'$ where $c = \sum_{i=1}^{n} c_i$

**Theorem 1** (Soundness): Given a file-getter $fg$ and a file $f$ if $\Gamma \vdash fg(f) : t(\vec{x})$ and a execution of the abstract machine $(fg(f) : nil, fg, [], [], nil, f : nil) \stackrel{c}{\Longrightarrow} (nil, fg, [], [], nil, nil)$ then $\exists \sigma \in \vec{x} \to \mathbb{N}$ such that $c \sqsubseteq \sigma(t(\vec{x}))$

## D.1 Statement

### D.1.1 Composition

$$(\text{T-Comp}) \frac{\Gamma_0 \vdash S_1 : t_1 \triangleright \Gamma_1 ... \Gamma_{n-1} \vdash S_n : t_n \triangleright \Gamma_n}{\Gamma_0 \vdash \vec{S} : t_1 + ... + t_n \triangleright \Gamma_n}$$

From the inductive hypothesis we know that $S_1, ..., S_n$ has the type $t_1, ..., t_n$ such that the cost of evaluating all the statements has a cost of $t_1 + ... + t_n$. With this, the abstract machine can execute the trace $S_1 \Rightarrow S_n$

$$\langle S :: ... :: S_n :: rest, fg, ls, es, vs, ss \rangle \stackrel{c_1}{\Longrightarrow} ... \stackrel{c_n}{\Longrightarrow} \langle rest, fg, ls, es, vs, ss \rangle$$

In the trace each statement is evaluated such the cost of each statement $S_1, ..., S_n$ is $c_1, ..., c_n$ such that $c_1 \sqsubseteq t_1, ..., c_n \sqsubseteq t_n$. Therefore it must hold that $c_1 + ... + c_n \sqsubseteq t_1 + ... + t_n$.

### D.1.2 While

$$(\text{T-While}) \frac{\Gamma_1 \vdash e : t_1 \qquad \Gamma_1 \vdash \vec{S} : t_2 \triangleright \Gamma_2}{\Gamma_1 \vdash \text{While } e \ \vec{S} : x \cdot (t_1 + t_2) + t_1 \triangleright \sqcup \{\Gamma_1, \Gamma_2\}}$$

We know from the inductive hypothesis that $e : t_1$ and $\vec{S} : t_2$ means that the trace of $e$ and $\vec{S}$ being evaluated will cost at most $t_1 + t_2$. With this, from the abstract machine, we have the transition sequence:

$$\langle \text{While } e \ \vec{S}, fg, ls, es, vs, ss \rangle \Longrightarrow \langle e :: \text{While}' \ e \ \vec{S} :: rest, fg, ls, es, vs, ss \rangle$$
$$\stackrel{c_1}{\Longrightarrow} \langle \text{While}' e \ \vec{S} :: rest, fg, ls, es, v :: vs, ss \rangle \quad \text{where } v \neq 0$$
$$\Longrightarrow \langle \vec{S} :: e :: \text{While}' e \ \vec{S} :: rest, fg, ls, es, vs, ss \rangle$$
$$\stackrel{c_2}{\Longrightarrow} \langle e :: \text{While}' e \ \vec{S} :: rest, fg, ls, es, vs, ss \rangle$$

where each iteration $i$ goes through the trace starting from the second transition.

When the condition costs $c_1$ such that $c_1 \sqsubseteq t_1$, and the statement costs $c_2$ such that $c_2 \sqsubseteq t_2$, and $i \leq x$, then it holds for all iterations that $i \cdot (c_1 + c_2) \sqsubseteq x \cdot (t_1 + t_2)$, where the while statement condition holds.

When the condition does not hold, only the condition is evaluated in the abstract machine, and we continue to the remaining program "*rest*".

$$\langle e :: \text{While}' \ e \ \vec{S} :: rest, fg, ls, es, vs, ss \rangle$$
$$\stackrel{c_1}{\Longrightarrow} \langle \text{While}' e \ \vec{S} :: rest, fg, ls, es, v :: vs, ss \rangle \quad \text{where } v = 0$$
$$\Longrightarrow \langle rest, fg, ls, es, vs, ss \rangle$$

Adding the final while statement condition check we get that $(i \cdot (c_1 + c_2) + c_1) \sqsubseteq (x \cdot (t_1 + t_2) + t_1)$ where $i \leq x$. Since we can assign the free variable $x$ to any value as long as it is equal or greater than $i$, the soundness theorem holds for the While statement. This also implies that the best estimate for $x$ is $i$.

### D.1.3 For loop

$$(\text{T-For}) \frac{\Gamma_1 \vdash \vec{S} : t_1 \triangleright \Gamma_2 \qquad n_1 \leq n_2}{\Gamma_1 \vdash \text{For } id \ n_1 \ n_2 \ \vec{S} : t_2 \cdot t_1 + t_2 \triangleright \Gamma_2} \quad t_2 = (n_2 - n_1 + 1)$$

We know from the inductive hypothesis that the for-loop body has a cost of $\vec{S} : t_1$ and runs $t_2 = n_2 - n_1 + 1$ iterations as per the side condition. With this, from the abstract machine, we have the transition sequence:

$$\langle \text{For } id\ n_1 n_2\ \vec{S} :: rest, fg\ ls, es, vs, ss \rangle$$
$$\implies \langle \text{Let } id\ n_1 :: \vec{S} :: ... :: \text{Let } id\ n_i :: \vec{S} :: ... :: \text{Let } id\ n_2 :: \vec{S} :: rest, fg, ls, es, vs, ss \rangle$$
$$\text{where } n_i = \mathbb{N}^{-1}(\mathbb{N}(n_1) + i) \text{ and } \mathbb{N}(n_1) \leq \mathbb{N}(n_2)$$

In the trace, the for-loop is unrolled and the current index in the loop is bound for each iteration starting at $n_1$ and ending at $n_2$ giving a total of $c_2 = n_2 - n_1 + 1$ iterations of the for loop. Therefore, it must hold that $c_2 \sqsubseteq t_2$.

The statement of the iteration has a cost of $c_1$ such that $c_1 \sqsubseteq t_1$. As the index is bound for each iteration and the cost of a let binding is 1, it must hold that an additional cost per iteration of $t_2$ must be added to account for the bindings. Therefore it must hold that $c_2 \cdot c_1 + c_2 \sqsubseteq t_2 \cdot t_1 + t_2$

### D.1.4 Component Call

$$(\text{T-CompCall})\ \frac{\begin{array}{c} \Gamma \vdash e_{2_1} : t_{11} \\ \vdots \\ \Gamma \vdash e_{2_n} : t_{nn} \\ \Gamma \vdash e_1 : t_1 \to ... \to t_n \to t_r \end{array} \quad \begin{cases} t_{ii} \sqsubseteq t_i \mid t_i \in \{t \to t, \{id:t\}\} \\ t_i \sqsubseteq 0 \mid \text{otherwise} \end{cases} \text{ where } i \in 1...n}{\Gamma \vdash \text{CompCall } e_1 \vec{e_2} : t_r + t_a + n \triangleright \Gamma\ t_a = \sum_{1...n} \begin{cases} 0 \mid t_{ii} \in \{t \to t, \{id:t\}\} \\ t_{ii} \mid \text{otherwise} \end{cases}}$$

From the inductive hypothesis we know that $e_1$ costs at most $t_r$ where $t_1$ to $t_n$ are bound to their concrete type. With this, the abstract machine can execute the transition, binding all the CompCall arguments. From the inductive hypothesis, we know that the statements in the closure of $e_1$ costs at most $t_r$, where $t_1$ to $t_n$ are bound to their concrete type. With this, the abstract machine can execute the transition, binding all the CompCall arguments.

$$\langle \text{CompCall } e_1 \vec{e_2} :: rest, fg, ls, es, vs, ss \rangle$$
$$\implies \langle e_1 :: \text{CompCall}'\ \vec{e_2} :: rest, fg, ls, es, vs, ss \rangle$$
$$\implies \langle \text{CompCall}'\ \vec{e_2} :: rest, fg\ ls, es, (\vec{id}, \vec{S}, s) :: vs, ss \rangle$$
$$\implies \langle e_1 :: ... :: e_n :: \text{PushScope } s :: \text{Bind } id_n :: ... :: \text{Bind } id_1 :: \vec{S} :: \text{PopScope} :: rest, fg, ls, es, vs, ss \rangle$$
$$\stackrel{c_1}{\implies} \langle \vec{S} :: \text{PopScope} :: rest, fg, ls, es, vs, s :: ss \rangle$$

The cost of binding the arguments is the cost of evaluating the expressions and the cost of binding the arguments denoted as $c_1$. The number of Bind's is $n$, where $n$ is the number of arguments from the CompCall typing judgement. Therefore since $c_1$ is the cost of both the binding and cost of the argument expressions then $c_1 \sqsubseteq t_a + n$ where $t_a$ is the cost of evaluating the argument expressions in the CompCall typing judgement. With this, the abstract machine can execute the transition of the components body.

$$\langle \vec{S} :: \text{PopScope} :: rest, fg, ls, es, vs, s :: ss \rangle$$
$$\stackrel{c_2}{\implies} \langle \text{PopScope} :: rest, fg, ls, es, vs, s :: ss \rangle$$
$$\implies \langle rest, fg, ls, es, vs, ss \rangle$$

The cost of the trace evaluating $\vec{S}$ is $c_2$ such that:
$$c_2 \sqsubseteq t_r \text{ given } \begin{cases} t_{ii} \sqsubseteq t_i & \mid t_i \in \{t \to t, \{id:t\}\} \\ t_i \sqsubseteq 0 & \mid \text{otherwise} \end{cases} \text{ where } i \in 1...n$$

Where the side condition ensures that cost of evaluating the expression is only counted once by setting the argument cost to 0, given that it is not an Object or a CompDef. Therefore it holds that $c_1 + c_2 \sqsubseteq t_r + t_a + n$.

### D.1.5 If

$$\text{(T-IF)} \frac{\begin{array}{c} \Gamma \vdash e : t_1 \\ \Gamma \vdash \vec{S_1} : t_2 \triangleright \Gamma_1 \\ \Gamma \vdash \vec{S_2} : t_3 \triangleright \Gamma_2 \end{array}}{\Gamma \vdash \text{If } e \ \vec{S_1} \ \vec{S_2} : t_1 + (t_2 \uparrow t_3) \triangleright \sqcup \{\Gamma_1, \Gamma_2\}}$$

We know from the inductive hypothesis that $e : t_1$ and $\vec{S_1} : t_2$ and $\vec{S_2} : t_3$ means that the trace of $e$ and either $\vec{S_1}$ or $\vec{S_2}$ being evaluated will cost at most $t_1 + (t_2 \uparrow t_3)$. With this, from the abstract machine, we can go from:

$$\langle \text{If } e_1 \vec{S_1} \vec{S_2} :: rest, fg, ls, es, vs, ss \rangle \Longrightarrow \langle e_1 :: \text{Branch } \vec{S_1} \vec{S_2} :: rest, fg, ls, es, vs, ss \rangle \stackrel{c_1}{\Longrightarrow}$$
$$\langle \text{Branch } \vec{S_1} \vec{S_2} :: rest, fg, ls, es, v :: vs, ss \rangle$$

Where $c_1$ is the cost of evaluating the condition, such that $c_1 \sqsubseteq t_1$.

Depending on whether the condition holds, there are two traces.

$$\langle \text{Branch } \vec{S_1} \vec{S_2} :: rest, fg, ls, es, v :: vs, ss \rangle \Longrightarrow \langle \vec{S_1} :: rest, fg, ls, es, vs, ss \rangle \text{ where } v \neq 0$$
$$\langle \text{Branch } \vec{S_1} \vec{S_2} :: rest, fg, ls, es, v :: vs, ss \rangle \Longrightarrow \langle \vec{S_2} :: rest, fg, ls, es, vs, ss \rangle \text{ where } v = 0$$

The cost of the trace evaluating $\vec{S_1}$ will be $c_2$ and the trace for $\vec{S_2}$ will be $c_3$ such that $c_2 \sqsubseteq t_2$ and $c_3 \sqsubseteq t_3$. Since it is not known which branch is evaluated, the worst case is selected such that $c_2 \sqsubseteq (t_2 \uparrow t_3)$ and $c_3 \sqsubseteq (t_2 \uparrow t_3)$. Adding the cost of the condition of the if statement we get that $c_1 + c_2 \sqsubseteq t_1 + (t_2 \uparrow t_3)$ and $c_1 + c_3 \sqsubseteq t_1 + (t_2 \uparrow t_3)$ therefore the soundness theorem holds for both branches of the (T-If) statement rule.

### D.1.6 Let Simple

$$\text{(T-LETSIMPLE)} \frac{\Gamma \vdash e : t_1}{\Gamma \vdash \text{Let } id \ e : 1 + t_1 \triangleright \Gamma[id \to 0]} t_1 \notin \{t_1 \to t_2, \{id : t_3\}\}$$

We know from the inductive hypothesis that $e : t_1$ means that the trace of $e$ being evaluated will use at most $t_1$ and any further use of the value will cost 0. With this, the abstract machine can execute the transition, binding $id$ to 0

$$\langle \text{Let } id \ e :: rest, fg, ls, es, vs, ss \rangle \Longrightarrow \langle e :: \text{Bind } id :: rest, fg, ls, es, vs, ss \rangle \stackrel{1}{\Longrightarrow}$$
$$\langle rest, fg, ls[(s, id) \to v], es, vs, s :: ss \rangle$$

Using $c$, where $c \sqsubseteq t_1$. We then use the Bind rule, which adds a cost of 1. Since we know that $c \sqsubseteq t_1$, we know it holds that $c + 1 \sqsubseteq t_1 + 1$.

### D.1.7 Let Advanced

$$\text{(T-LETADVANCED)} \frac{\Gamma \vdash e : t_1}{\Gamma \vdash \text{Let } id \ e : 1 \triangleright \Gamma[id \to t_1]} t_1 \in \{t_1 \to t_2, \{id : t_3\}\}$$

We know that $t_1$ is a CompDef that cost 0 until called, or a variable, which also cost 0. With this, the abstract machine can execute the transition, binding $id$ to $t_1$

$$\langle \text{Let } id\ e :: rest, fg, ls, es, vs, ss \rangle \implies \langle e :: \text{Bind } id :: rest, fg, ls, es, vs, ss \rangle \xRightarrow{1}$$
$$\langle rest, fg, ls[(s, id) \to v], es, vs, s :: ss \rangle$$

We know that binding a variable costs exactly $c = 1$ which is equivalent to the estimated cost $t = 1$ therefore it holds that $c \sqsubseteq t$.

### D.1.8 Assign Simple

$$(\text{T-AssignSimple}) \frac{\Gamma \vdash e : t_1 \qquad id \in \text{dom}(\Gamma)}{\Gamma \vdash \text{Assign } id\ e : 1 + t_1 \rhd \Gamma[id \to 0]} \quad t_1 \notin \{t_1 \to t_2, \{id : t_3\}\}$$

We know from the inductive hypothesis that $e : t_1$ means that the trace of $e$ being evaluated will cost at most $t_1$ and any further use of the value will cost 0. With this, the abstract machine can execute the transition, binding $id$ to 0:

$$\langle \text{Assign } id\ e :: rest, fg, ls, es, vs, ss \rangle \implies \langle e :: \text{Bind } id :: rest, fg, ls, es, vs, ss \rangle \xRightarrow{1}$$
$$\langle rest, fg, ls[(s, id) \to v], es, vs, s :: ss \rangle$$

Using the cost $c$, where $c \sqsubseteq t_1$. We then use the (R-Bind) rule, which adds a cost of 1. Since we know that $c \sqsubseteq t_1$, we know it holds that $c + 1 \sqsubseteq t_1 + 1$.

### D.1.9 Assign Advanced

$$(\text{T-AssignAdvanced}) \frac{\Gamma \vdash e : t_1 \qquad id \in \text{dom}(\Gamma)}{\Gamma \vdash \text{Assign } id\ e : 1 \rhd \Gamma[id \to t_1]} \quad t_1 \in \{t_1 \to t_2, \{id : t_3\}\}$$

We know that $t_1$ is a CompDef that cost 0 until called. With this, the abstract machine can follow the trace, binding $id$ to $t_1$

$$\langle \text{Let } id\ e :: rest, fg, ls, es, vs, ss \rangle \implies \langle e :: \text{Bind } id :: rest, fg, ls, es, vs, ss \rangle \xRightarrow{1}$$
$$\langle rest, fg, ls[(s, id) \to v], es, vs, s :: ss \rangle$$

We know that binding a variable costs exactly $c = 1$ which is equivalent to the estimated cost $t = 1$ therefore it holds that $c \sqsubseteq t$.

### D.1.10 Source File

$$(\text{T-SrcFile}) \frac{\begin{array}{c} \Gamma_0 \vdash im_1 : t_{1_1} \rhd \Gamma^1 ... \Gamma^{n-1} \vdash im_n : t_{1_n} \rhd \Gamma^n \\ \Gamma^n \vdash \vec{S} : t_{2_1} + ... + t_{2_m} \rhd \Gamma^{n+m} \\ \Gamma^{n+m} \vdash ex_1 : t_{3_1} \rhd \Gamma^{n+m+1} ... \Gamma^{n+m+l-1} \vdash ex_l : t_{3_l} \rhd \Gamma^{n+m+l} \end{array}}{\Gamma_0 \vdash \text{SrcFile } \overrightarrow{im}\ \vec{S}\ \vec{ex} : t_{1_1} + ... + t_{1_n} + t_{2_1} + ... + t_{2_m} + t_{3_1} + ... + t_{3_l} \rhd \Gamma^{n+m+l}}$$

We know that from the inductive hypothesis that the cost of the imports, statements and exports are $t_{im} = t_{1_1} + ...t_{1_n}$, $t_S = t_{2_1} + ...t_{2_m}$ and $t_{ex} = t_{3_1} + ... + t_{3_l}$. With this, from the abstract machine, we have the transition sequence:

$$\langle \overrightarrow{\text{Import}}\ \vec{S}\ \overrightarrow{\text{Export}} :: rest, fg, ls, es, vs, ss \rangle \implies \langle \text{Import}_1 :: ... :: \text{Import}_n :: S_1 :: ... :: S_m ::$$
$$\text{Export}_1 :: ... :: \text{Export}_j :: rest, fg, ls, es, vs, ss \rangle$$

Where the sub trace that evaluates all the imports is $c_1$ such that $c_1 \sqsubseteq t_{im}$, and the trace that evaluates all the statements is $c_2$ such that $c_2 \sqsubseteq t_S$. The final sub trace of exports is $c_3$ such that $c_3 \sqsubseteq t_{ex}$ therefore it must hold that $c_1 + c_2 + c_3 \sqsubseteq t_{im} + t_S + t_{ex}$.

## D.2 Imports

### D.2.1 Import-All

$$(\text{T-ImportAll}) \frac{[] \vdash fg(\text{File}) : t \triangleright \Gamma'}{\Gamma \vdash \text{Import } id \text{ File} : t + 3 \triangleright \Gamma[id \to \Gamma'(\varepsilon), \varepsilon \to \{\}]}$$

From the inductive hypothesis we know that $fg(\text{file})$ will take at most $t$. If we look at the abstract machines trace for (R-ImportAll) we have:

$$\langle id\ f :: rest, fg, ls, es, vs, ss \rangle \stackrel{2}{\Longrightarrow} \langle fg(f) :: \text{PopScope} :: \text{BindAll } id :: \text{EmptyExports} :: rest, fg, ls, es, vs, f :: ss \rangle$$

Giving us an initial cost of 2. We know that $fg(f)$ produce a new list of instructions that will cost $c$ to evaluate such that $c \sqsubseteq t$. The final part of the trace we have BindAll, which has the constant cost of 1.

$$\langle \text{PopScope} :: rest, fg, ls, es, vs, s :: ss \rangle$$
$$\Longrightarrow \langle \text{BindAll } id :: \text{EmptyExports} :: rest, fg, ls, es, vs, s :: ss \rangle$$
$$\stackrel{1}{\Longrightarrow} \langle \text{EmptyExports} :: rest, fg, ls[(s, id) \to es], es, vs, s :: ss \rangle$$
$$\Longrightarrow \langle rest, fg, ls, [], vs, ss \rangle$$

Therefore, for the (R-ImportAll) rule it should hold that $c + 1 + 2 \sqsubseteq t + 3$.

### D.2.2 Import Selected

$$(\text{T-ImportSelected}) \frac{[] \vdash fg(\text{File}) : t \triangleright \Gamma'\left[\varepsilon' \to \{id_i : t_i\}_{i \in \{1..n..m\}}\right]}{\Gamma \vdash \text{Import } \vec{id}^n \text{ File} : t + n + 2 \triangleright \Gamma[id_1 \to t_1, ..., id_n \to t_n, \varepsilon \to \{\}]}$$

From the inductive hypothesis we know that $fg(\text{File})$ will cost at most $t$. If we look at the trace of the abstract machine, for (R-ImportSelected), we have:

$$\langle \vec{id}\ f :: rest, fg, ls, es, vs, ss \rangle \stackrel{2}{\Longrightarrow} \langle fg(f) :: \text{PopScope} :: \overrightarrow{\text{BindSelected } id} :: \text{EmptyExports} :: rest, fg, ls, es, vs, f :: ss \rangle$$

Giving us an initial cost of 2. We know that when $fg(f)$ is expanded, it results in a sequence of instructions. The execution of this transition sequence will have a cost of $c$ such that $c \sqsubseteq t$. Then, if we look at the next step of the trace, we have that pop-scope costs 0.

$$\langle \text{PopScope} :: \overrightarrow{\text{BindSelected } id} :: \text{EmptyExports} :: rest, fg, ls, es, vs, f :: ss \rangle \Longrightarrow$$
$$\langle \overrightarrow{\text{BindSelected } id} :: \text{EmptyExports} :: rest, fg, ls, es, vs, ss \rangle$$

We can expand the trace into $\langle \text{BindSelected } id :: ... :: \text{BindSelected} :: \text{EmptyExports} :: rest, fg, ls, es, vs, ss \rangle$ where the number of BindSelected corresponds to $m = |\overrightarrow{\text{BindSelected } id}|$. The trace for a single BindSelected in the abstract machine is as follows:

$$\langle \text{BindSelected } id :: rest, fg, ls, es, vs, s :: ss \rangle \stackrel{1}{\Longrightarrow} \langle rest, fg, ls[(s, id) \mapsto es(id)], es, vs, s :: ss \rangle$$

We observe that the cost of each of the $m$ BindSelected is 1, and as $m$ is equivalent to the number of $id$'s, it must hold that $m \sqsubseteq n$. We can observe that the final trace of the transition sequence until "$rest$" has a cost of 0.

$$\langle \text{EmptyExports} :: rest, fg, ls, es, vs, ss \rangle \Longrightarrow \langle rest, fg, ls, [], vs, ss \rangle$$

Since $m \sqsubseteq n$ and $c \sqsubseteq t$, it must hold that $c + m \sqsubseteq t + n$. With the initial cost of 2 in the abstract machine, which can be observed in the type judgement for the Type Rule 6 rule, it must hold that $c + m + 2 \sqsubseteq t + n + 2$.

## D.3 Export

$$(\text{T-Export}) \frac{\begin{array}{c} t = \Gamma(id) \\ \{id_i : t_i\}_{i \in \{1..n\}} = \Gamma(\varepsilon) \end{array}}{\Gamma \vdash \text{Export } id : 1 \triangleright \Gamma[\varepsilon \to ex] \quad ex = \{id_1 : t_1, ..., id_n : t_n, id : t\}_{i \in 1...n+1}}$$

The Export rule always gets typed as the constant $t = 1$. If we look at the trace of the abstract machine for export we have:

$$\langle \text{Export } id :: rest, fg, ls, es, vs, s :: ss \rangle \implies \langle rest, fg, ls, es[id \to ls(s, id)], vs, s :: ss \rangle$$

Since the trace of the export is constant $c = 1$, it must hold that $c \sqsubseteq t$.

## D.4 Expressions

### D.4.1 Num

$$(\text{T-Num}) \frac{}{\Gamma \vdash n : 0}$$

The (T-Num) rule is typed with the constant $t = 0$. When examining the abstract machine trace for (R-Num), we observe:

$$\langle n :: rest, fg, ls, es, vs, ss \rangle \implies \langle rest, fg, ls, es, v :: vs, ss \rangle \text{ where } v = \mathbb{N}(n)$$

that the transition in the abstract machine costs $c = 0$ and $c \sqsubseteq t$.

### D.4.2 Var

$$(\text{T-Var}) \frac{}{\Gamma \vdash id : t} t = \Gamma(id)$$

From the side condition we know that the type of the variable is $t$. When examining the abstract machine trace for (R-Var), we observe:

$$\langle \text{Var } id :: rest, fg, ls, es, vs, s :: ss \rangle \implies \langle rest, fg, ls, es, ls(s, id) :: vs, s :: ss \rangle$$

that the trace has a cost of $c = 0$ using that 0 is the least element it must hold that for any $t$ that $0 \sqsubseteq t$

### D.4.3 BinOp

$$(\text{T-BinOp}) \frac{\Gamma_1 \vdash e : t_1 \quad \Gamma_1 \vdash e : t_2 \quad t_3 = bc(\text{BinOp})}{\Gamma \vdash \text{BinOp } e_1 e_2 : t_1 + t_2 + t_3} t_1, t_2 \notin \{t \to t, \{id : t\}\}$$

The BinOp rule is typed with the cost of $t_1 + t_2 + t_3$. From the inductive hypothesis, we know that $e_1$ can be reduced with a cost of at most $t_1$, and $e_2$ can be reduced with a cost of at most $t_2$. We also know that $t_3$ is equal to $bc(\text{BinOp})$.

$$\langle \text{BinOp } e_1\ e_2 :: rest, fg, ls, es, vs, ss \rangle \underset{c_2}{\overset{c_1}{\implies}} \langle e_1 :: e_2 :: \text{BinOp} :: rest, fg, ls, es, vs, ss \rangle \overset{c_1}{\underset{c_3}{\implies}}$$
$$\langle e_2 :: \text{BinOp} :: rest, fg, ls, es, v_1 :: vs, ss \rangle \overset{c_2}{\implies} (\text{BinOp} :: rest, fg, ls, es, v_2 :: v_1 :: vs, ss) \implies$$
$$(rest, fg, ls, es, v_3 :: vs, ss) \text{ where } v_3 = \text{BinOp}(v_1, v_2) \text{ and } c_3 = bc(\text{BinOp})$$

The execution of the two sub expressions has a cost of $c_1 + c_2$ such that $c_1 + c_2 \sqsubseteq t_1 + t_2$ and the BinOp has a cost of $c_3$ such that $c_3 \sqsubseteq t_3$. Therefore it must hold that $c_1 + c_2 + c_3 \sqsubseteq t_1 + t_2 + t_3$.

### D.4.4 Projection

$$(\text{T-Proj}) \frac{\Gamma(id_1) = \{l_i : t_i\}_{i \in I}}{\Gamma \vdash \mathsf{Proj}\ id_1 id_j : t_j} \text{ where } j \in I$$

From the side condition we know that the type of the variable is $t_i$. When examining the abstract machine trace for (R-Proj), we observe:

$$\langle \mathsf{Proj}\ id_1 id_2 :: rest, fg, ls, es, vs, s :: ss \rangle \implies \langle rest, fg, ls, es, \mathsf{Obj}(id_2) :: vs, s :: ss \rangle$$
$$\text{where } \mathsf{Obj} = ls(s, id_1)$$
$$\text{and } \mathsf{Obj} :: (id \to v)$$

that the trace has a cost of $c = 0$ using that 0 is the least element it must hold that for any $t_i$ that $0 \sqsubseteq t_i$

### D.4.5 Component Definition

$$(\text{T-ComDef}) \frac{\Gamma[id_1 : t_1, ..., id_n : t_n] \vdash \vec{S} : t_r}{\Gamma \vdash \mathsf{CompDef}\ \vec{id}\ \vec{S} : t_1 \to ... \to t_n \to t_r}$$

The CompDef rule is typed as an arrow type $t_1 \to ... \to t_n \to t_r$ where $t_1$ to $t_n$ are free variables corresponding to the arguments to the component. When examining the abstract machine trace for CompDef, we observe:

$$\langle \mathsf{CompDef}\ \vec{id}\ \vec{S} :: rest, fg, ls, es, vs, s :: ss \rangle \implies \langle rest, fg, ls, es, \left(\vec{id}, \vec{S}, s\right) :: vs, s :: ss \rangle$$

The cost of creating a component is $c = 0$, and therefore it must hold that $c \sqsubseteq t$.

# E Abstract Machine Trace

(R-SrcFile)(R-ImportSelected)(R-SrcFile)(R-Let)(R-Num)
(R-Bind)(R-While)(R-Var)(R-WhileTrue)(R-Assign)
(R-BinOp1)(R-Var)(R-Num)(R-BinOp2)(R-Bind)
(R-Var)(R-WhileTrue)(R-Assign)(R-BinOp1)(R-Var)
(R-Num)(R-BinOp2)(R-Bind)(R-Var)(R-WhileTrue)
(R-Assign)(R-BinOp1)(R-Var)(R-Num)(R-BinOp2)
(R-Bind)(R-Var)(R-WhileFalse)(R-Export)(R-PopScope)
(R-BindSelected)(R-EmptyExports)(R-Let)(R-Num)(R-Bind)
(R-Let)(R-CompDef)(R-Bind)(R-CompCall)(R-Var)
(R-CompCallPrime)(R-Var)(R-PushScope)(R-Bind)(R-Assign)
(R-BinOp1)(R-Var)(R-Num)(R-BinOp2)(R-Bind)
(R-PopScope)

Figure 6: Trace for the example on Listing 2

From the trace, we get that main.jsx imports a variable using (R-ImportSelected), and we therefore run the entire file containing the variable (simpleWhile.jsx), which ends with the variable x being put on the exports stack by (R-Export), and (R-PopScope) being used to exit out of the file. When this is completed, the remainder of main.jsx is executed.